\documentclass{eas}
\usepackage{graphicx}
\begin{document}

\title{Formation of low-mass stars and brown dwarfs} 
\author{Patrick Hennebelle}\address{Laboratoire de radioastronomie, Ecole normale sup\'erieure and Observatoire de Paris,  \\ UMR CNRS 8112 24 rue Lhomond, 75231 Paris Cedex 05, France  \\ email: {\tt patrick.hennebelle@ens.fr}}
%
%
\begin{abstract}
These lectures attempt to expose the most important ideas, 
which have been proposed to explain the formation of stars
with particular emphasis on the formation of brown dwarfs 
and low-mass stars. We first describe the important 
physical processes which trigger the collapse of a self-gravitating 
piece of fluid and  regulate the star formation rate in molecular 
clouds. Then  we review the various theories which have been proposed 
along the years
to explain the origin of the stellar initial mass function paying particular 
attention to  four models, namely  the competitive accretion and the 
theories based respectively  on stopped accretion, MHD shocks and 
turbulent dispersion. As it is yet unsettled whether the 
brown dwarfs form as low-mass stars, we present the theory 
of brown dwarfs based on disk fragmentation stressing all the 
uncertainties due to the radiative feedback and magnetic field. 
Finally, we describe the results of large scale simulations
performed to explain the collapse and fragmentation of molecular clouds.
\end{abstract}
\maketitle
\section{Introduction}
Stars are the building blocks of our universe and understanding 
their formation and evolution is one of the most important problems 
of astrophysics. It is well established that stars form by gravitational collapse of molecular 
dense cores. These cores are themselves embedded in molecular clouds
and often, though not always, inside filaments. However, the conditions 
through which the dense cores form, collapse and fragment remain a matter of debate. 
Three questions are of particular importance: the star formation rate (hereafter SFR), the problem 
of the initial mass function (hereafter IMF), and the problem of disk formation and evolution that 
may give birth  to brown dwarfs, low-mass stars and planets.

As pointed out by Zuckerman \& Evans (1974), 
if all the molecular gas observed in the Galaxy was collapsing in a 
freefall time, then the star formation rate would be 
10 to 100 times higher than the 
star formation rate of 
$\simeq 3$ M$_\odot$ yr$^{-1}$ observed in the Galaxy.
A lot of efforts have been devoted  to  explain this 
low efficiency of  star formation. Two main ideas have been 
explored so far. The first is that magnetic field provides 
an efficient support against gravity and delays the star 
formation (e.g. Shu et al. 1987). The second postulates
that turbulent motions observed in molecular clouds, prevent the clouds
to collapse in a freefall time (Mac Low \& Klessen 2004, McKee \& Ostriker 2007). 

Of particular importance is the problem 
of the initial mass function of stars (Salpeter 1955, Scalo 1986,
Kroupa 2002, Chabrier 2003) largely 
because the star properties, evolution and influence on the surrounding interstellar medium
  strongly depend on their masses. 
 It is generally found that 
the number of stars per logarithmic bin of masses,
$d N / d \log M$,  can be described
by a lognormal 
distribution below 1 $M_\odot$,  peaking at about $\simeq$0.3 $M_\odot$, and a power-law of slope 
$-1.3$ for masses between 1 and 10 $M_\odot$\footnote{ It should be kept in mind that the {\it initial} mass function is measured only up to
about 8 $M_\odot$, in young stellar clusters, and is inferred only indirectly for larger masses (see e.g. Kroupa 2002)} (e.g. Chabrier 2003).
It should be stressed that the IMF of more massive stars is extremely 
poorly known.

The conservation of angular momentum during collapse leading to the so-called
centrifugal barrier is another long-standing problem in astrophysics. Angular momentum
is responsible for the formation of ubiquitous circumstellar disks 
(e.g. Haisch et al. 2001, Watson et al. 2007) and probably, at least in part for the formation 
of binaries (Duquennoy \& Mayor 1991). Indeed, the fragmentation of 
massive early, class-0, disks is possibly a mode for the formation of low-mass 
stars and brown dwarfs
while the formation of planets  occurs within disks probably at different 
stages of their evolution.

In the following, we first present the basic theory of self-gravitating 
isothermal gas, including Jeans length and mass, self-gravitating equilibrium and 
collapse. We also infer the mass of the smallest star that could possibly form 
in present day universe. We then discuss the question of the star formation rate, presenting
the two main schools of thought which have been proposed so far, 
namely the magnetic  and the turbulent regulation of star formation. Note that 
as this aspect does not constitute the main goal of this lecture but 
is presented for completeness, only the most important aspects are discussed.
The third section is devoted to the question of the initial mass function and 
the various theories, which have been proposed along the years. Four different
theories will be presented, namely the competitive accretion scenario, the theories
based on stopped accretion, the 
MHD shock based theory, and the turbulent dispersion theory.
In the fourth section, we first discuss self-gravitating disk formation, 
evolution and stability. Then we present the brown dwarf formation by 
disk fragmentation theory stressing the  
uncertainties due to the radiative feedback and magnetic field. 
Finally, in the fifth section, we describe the results of large scale 
numerical simulations which have been performed so far to study the formation 
of stars in massive collapsing clumps. In particular, 
we present two other scenarii for the formation of brown dwarfs which have 
been studied in these simulations, namely the formation by ejection and the 
formation in collapsing filaments. Then we discuss the IMF which have 
been obtained in these simulations as well as the physical explanations
that have been proposed.

\section{Physical processes}

\subsection{Gravity and thermal support}
Before going into the more complex situations of magnetized and 
turbulent clouds, we start by establishing the  basic principles 
of self-gravitating isothermal gas dynamics.

\subsubsection{Ratio of thermal and gravitational energies}
It is instructive to start by computing the ratio between
the thermal energy, 
$E_{\rm therm}= {M \over (\gamma -1) m_p} k_b T$ and the 
gravitational energy $E_{\rm grav} = (3/5) M^2 G / R $, where 
$M$ is the cloud mass, $R$ its radius, $m_p$ the mean mass
per particle, $T$ the temperature, $k_b$ and G are respectively the Boltzmann and the gravitational constants, 
and $\gamma$ is the adiabatic index, which depends on the internal 
degrees of freedom of the constituents. 
For a polytropic cloud, the thermal pressure is given by 
$P = K \rho^{\Gamma}$, where $\Gamma$ is an effective adiabatic 
exponent that depends on the cooling processes. With these expressions, we get
\begin{eqnarray}
{ E_{\rm therm} \over E_{\rm grav}} \propto R^{4-3 \Gamma}.
\label{ratio_ener}
\end{eqnarray}
This expression clearly shows that $\Gamma =4/3$ is a critical case below
which thermal pressure is unable to support the cloud against gravitational 
collapse because the ratio between support and gravitational energy 
drops with the radius. While this is in particular true for the  isothermal case,
$\Gamma=1$, gravitational collapse will be stopped by thermal pressure if the gas 
is unable to cool efficiently both for a monoatomic gas ($\Gamma \simeq \gamma = 5/3$)
and for a diatomic one  ($\Gamma \simeq \gamma = 7/5$).

\subsubsection{Jeans length, Jeans mass, and freefall time}
The Jeans length (Jeans 1905, Lequeux 2005) is easily derived by performing a linear 
analysis of the self-gravitating fluid equations.
Let us consider a cloud of density $\rho_0$, radius $R$, and sound speed $C_s$ (note that
strictly speaking a self-gravitating isothermal cloud cannot have a uniform density because the pressure
forces should compensate the gravitational forces).
A linear analysis leads to the dispersion relation
\begin{eqnarray}
\omega^2 = C_s^2 k ^2 - 4 \pi G \rho_0, 
\label{eq1}
\end{eqnarray} 
which reveals that when the wave number, $k$, is smaller than 
$\sqrt{4 \pi G \rho_0} / C_s$, the waves cannot propagate and
perturbations are amplified. From this we obtain the Jeans length, $\lambda_J$,
\begin{eqnarray}
\lambda_J =  \sqrt{\pi C_s^2 \over G \rho_0 },
\end{eqnarray}
where $G$ is the gravitational constant. 
The Jeans length can be physically understood in the following way.
 Self-gravity tends to induce contraction in a time scale
of the order of $1/\sqrt{G \rho_0}$. On the other hand, thermal pressure tends to reestablish 
uniform density in a sound  crossing time, $R / C_s$.
If  $1/\sqrt{G \rho_0} < R / C_s$, then the waves cannot erase the pressure fluctuations induced 
by the gravitational contraction before the whole cloud collapses. 

The Jeans mass is naturally defined as the mass contained in a volume 
of typical size $\lambda_J$. 
The Jeans mass is generally determined as 
\begin{eqnarray}
\nonumber
M_J &=& 4 \pi /3 \rho _0 (\lambda_J/2)^3 \\
&=& {\pi ^{5/2} \over 6}  {C_s^{3} \over \left( G ^3 \rho_0  \right)^{1/2} },
\label{jeans_mass}
\end{eqnarray}
though there is no fundamental justification for this choice within a factor of a few.

Equation~\ref{jeans_mass} shows that, for an isothermal gas, the Jeans mass
 decreases with density. 
Therefore,  in a collapsing cloud the 
Jeans mass value increases as the collapse proceeds. 
 Based on this argument, Hoyle (1953) proposed the concept of recursive fragmentation by  which a cloud
is fragmenting more and more as it becomes denser. However, as shown by eq.~(\ref{eq1}), the 
growth rate of the gravitational instability, decreases with $k$, meaning that the large scale
perturbations evolve more rapidly than the smaller scale perturbations.
Thus the recursive fragmentation scenario suffers a timescale problem.
Note that there is an inconsistency in inferring eq.~(\ref{eq1}) as a uniform medium is not an exact 
solution of the self-gravitating fluid equations (a force should balance gravity). When 
exact solutions  of the equations (see next section) are perturbed (e.g. Nagai et al. 1998 for  layers or 
Fiege \& Pudritz 2000 for filaments), it is generally obtained that the growth rate
tends to zero as $k \rightarrow 0$ while the fastest growing mode corresponds to  a few 
times the Jeans length. 

In  general, it is not possible to  analytically compute the time for a cloud to collapse. 
However, in the ideal case of a cold spherical cloud with uniform density, one can calculate it exactly
(see e.g. Lequeux 2005). The result, known as the freefall time, is 
\begin{eqnarray}
\tau _{\rm ff} = \sqrt{ {3 \pi \over 32 G \rho_0 }}.
\end{eqnarray}

\subsubsection{The smallest Jeans mass in contemporary molecular clouds}
The expression stated by eq.~(\ref{jeans_mass}) assumes that the gas
follows a barotropic equation of state and neglects cooling and heating.
This is only valid above a certain limit that eventually sets the 
value of the smallest Jeans mass. 
To estimate its value we closely follow the approach of  Rees (1976) and Whitworth et al. (2007). 
The first condition that must be fulfilled is that the size of the piece of fluid, $R$, must be of 
the order of the Jeans length as explained above. Thus $R \simeq \lambda_J$, leading to
\begin{eqnarray}
R \simeq {6 \over \pi ^2} {G \over c_s^2} M _J.
\label{cond1}
\end{eqnarray}
The second condition is that the thermal energy due to the gravitational contraction, should 
be efficiently radiated. If this condition is not fulfilled, as explained above, the 
effective adiabatic index, $\Gamma$, will be larger than $4/3$ and thermal pressure will halt
the collapse.
The heating rate is given by the work of the thermal pressure $P d_t V$. Assuming that the collapse
is near freefall, we get $v = d_t R \simeq \sqrt{GM/R}$ and thus
\begin{eqnarray}
P d_t V = \rho c_s^2 {d \over dt} \left( {4 \pi \over 3} R^3 \right) \simeq {3 c_s^2 M \over R} \sqrt{GM \over R}.
\end{eqnarray}
The cooling is due to the radiative loss which in the optically thick regime is given by
(e.g. Mihalas \& Mihalas 1984)
\begin{eqnarray}
{\cal L} = {16 \pi  \over 3 \chi_R} R^2 \sigma T^3 \partial_R T
\end{eqnarray}
where $\chi _R=\rho \kappa$ and $\kappa$ is the opacity (e.g. Semenov et al. 2003) while
 $\sigma = 2 \pi^5 k_b^4 / 15 h^3 c^2$ the Stefan-Boltzmann constant.
Thus the second condition that the piece of gas must fulfill to be gravitationally unstable is 
given by
\begin{eqnarray}
P d_t V  \simeq {3 c_s^2 M \over R} \sqrt{GM \over R} \simeq {\cal L} \simeq {16 \pi  \over 3 \tau } R^2 \sigma T^4, 
\end{eqnarray}
where the optical depth $\tau \simeq \chi _R  R$.
This leads to 
\begin{eqnarray}
R  \simeq \left( {3^6 5^2 \over 2^{10} \pi^{12}} \right)^{1/7} \left( {G^2 h^6 c^4 \over c_s^{12} m_p^8 } \right)^{1/7}
\tau ^{2/7} M^{3/7}.
\label{cond2}
\end{eqnarray}
Combining the conditions stated by eqs.~(\ref{cond1})-(\ref{cond2}), one gets a characteristic mass 
given by 
\begin{eqnarray}
M \simeq \left(5^2 \pi^2 \over 2^{17} 3 \right)^{1/4} \left( {h c \over G}\right)^{3/2}
m_p^{-2} \left( {c_s \over c} \right)^{1/2} \tau^{1/2} \simeq
{m_{Planck}^3 \over m_p^2} \left( {c_s \over c} \right)^{1/2} \tau^{1/2}, 
\end{eqnarray}
where $m_{Planck}$ is the Planck mass. The exact value of the numerical factor 
depends on the  assumptions that have been made
  and may vary from one author to an other. 
One finds, for an optical depth $\tau \simeq 1$, that it is of the order 
of a few Jupiter masses. Note that as discussed by Whitworth \& Stamatellos (2006)
and Masunaga \& Inutsuka (2000), $\tau \ge 1$ is not a necessary condition although 
it appears reasonable in this context.

\subsubsection{Equilibrium configurations}
Equilibrium configurations are obtained when pressure forces compensate 
gravitational forces. Such static solutions of the fluid equations are useful and convenient guides.
They allow  to test numerical codes,  perform more rigorous stability analysis than the Jeans analysis
and can sometimes be compared directly to the observations.
The equations of equilibrium, namely the hydrostatic equation and the Poisson equation, are respectively:
\begin{equation}
-C_s^2 \partial _X \rho + \rho \partial_X \phi = 0,  
\label{eq2}
\end{equation}
\begin{equation}
{1 \over X ^{D-1}} \partial _X (X^{D-1} \partial_X \phi)  = - 4 \pi G \rho.
\label{eq3}
\end{equation}
Combining these 2 equations  leads to the so-called Lane-Emden equation: 
\begin{equation}
{1 \over X ^{D-1}} \partial _X \left(X^{D-1} {\partial_X \rho \over \rho} \right)  
= - {4 \pi G \over C_s^2} \rho, 
\label{eq4}
\end{equation}
where $D$ is the dimension and $X$ the spatial coordinate.  

In plane-parallel geometry ($D=1$), $X$ represents  the usual Cartesian coordinate, $z$, 
 whereas in cylindrical geometry ($D=2$), $X$ represents  the cylindrical radius, $r$. 
 In the first case, a self-gravitating layer solution has been inferred by Spitzer (1942)
  whereas in the second, a self-gravitating filament has been obtained by 
Ostriker (1964).  These two solutions are fully analytical. They are characterized by a flat 
density profile near $X=0$. The former presents an exponential decrease for 
large $z$ whereas the latter decreases as $r^{-4}$.

In spherical geometry ($D=3$) ($X$ represents the spherical radius, $r$),  
the solutions of eq.~(\ref{eq4}) are the so-called 
Bonnor-Ebert spheres (Bonnor 1956).
In general, these solutions are not analytical and must be obtained by solving numerically 
eq.~(\ref{eq4}). There is however a noticeable exception which is the singular isothermal 
sphere whose density  is given by $\rho_{\rm SIS} = C_s^2 / (2 \pi G r^2) $. 
The density profile of the Bonnor-Ebert  sphere is flat  in the inner part and 
tends toward the density of the singular isothermal sphere
 at large radii. 
Since it is physically required that the cloud has a 
finite radius, the solutions are obtained by truncating this profile at any arbitrary radius, assuming 
pressure equilibrium with the medium outside the cloud which is 
 supposed to be diffuse and warm.
Therefore a whole family of equilibrium solutions is obtained. They can be characterized
by the density contrast between the center and the edge. Stability analysis reveals that
the solutions which have a density contrast smaller than about 14 are stable and unstable 
otherwise.  

As mentioned in the previous section, 
stability analysis of the self-gravitating layer and filament have been performed 
in various studies (e.g. Larson 1985, Fiege \& Pudritz 2000). Both are unstable to 
perturbations of wavelengths 
comparable to the Jeans length. In particular, this suggests that cores or filaments 
distributed periodically, 
could form through
 gravitational  instability within self-gravitating filaments and layers 
respectively. As the interstellar medium is out of equilibrium and subject to 
supersonic motions, it is always difficult to assess this scenario quantitatively 
but qualitatively at least, spatially roughly periodically distributed cores and filaments 
are often observed (Dutrey et al. 1991).

\subsubsection{Gravitational collapse}
The gravitational collapse of a spherical cloud has been investigated in some details both 
analytically and numerically. 
Since even in spherical geometry, the cloud is described by  two non linear 
equations, the analytical models have mainly focused on self-similar solutions 
(e.g. Larson 1969; Penston 1969; Shu 1977; Whitworth \& Summers 1985) which allow to reduce the 
equations of the problem to simpler ordinary equations.
These solutions have been useful to understand the physics of the collapse and sometimes 
used in various contexts to provide easily time-dependent density and velocity fields.
Two main types of solutions have been inferred. Larson (1969) and Penston (1969)
derive a solution which presents supersonic infall velocity at large radii ($\simeq 3.3 C_s$)
whereas Shu (1977) obtains a solution in which the gas is initially at rest and 
undergoes inside-out collapse. A rarefaction wave which propagates outwards at the sound speed, 
is launched from the cloud center when the protostar  forms.  
All self-similar solutions  have  constant accretion rate equal to few up to several times 
$C_s^3 / G$. Note that in all solutions the density field is proportional to $r^{-2}$ in the 
outer part and to $r^{-3/2}$ in the inner region which has been reached by the rarefaction wave. 
Finally, note that the density of the Larson-Penston solution is about 8 times 
 larger than the density of the Shu's solution at infinity.

The collapse has also been investigated numerically. Larson (1969) starts with a uniform density 
cloud and calculates the gas contraction up to the formation of the protostar by using some 
simplified radiative transfer (see also Masunaga \& Inutsuka 2000). 
He shows that a first accretion shock develops at the edge of 
 the thermally supported core which forms  when the dust becomes
opaque to its own radiation, i.e. at a density of about $10^{-13}$ g cm$^{-3}$. This  core 
 is sometimes called the first Larson core. A second accretion shock forms at the edge 
of the protostar at much higher density ($\simeq 10^{-2}$ g cm$^{-3}$). 
 Foster \& Chevalier (1993) start with a slightly unstable 
Bonnor-Ebert sphere. Interestingly, they find that the collapse occurs very slowly in the 
outer part where only subsonic infall velocities develop. In the inner part, however, 
supersonic motions appear. Indeed, they show that convergence towards the Larson-Penston solution
is achieved deep inside the cloud. On the other hand, in the outer part of the envelope, 
the density profile turns out to stay close to $\rho_{\rm SIS}$.
Triggered collapse has also been investigated by various authors (e.g. Hennebelle et al. 2003). 
Faster infall velocities
are then obtained as well as densities few times denser than $\rho_{\rm SIS}$.
A common feature shared by the numerical solutions is that the accretion rate varies 
significantly along time, unlike what is inferred from the self-similar solutions.

\subsection{Influence of  magnetic field}
As recalled previously, magnetic field has early been  proposed to provide 
an important mechanical support to the gas (see e.g. Shu et al. 1987)
which could possibly explain the low star formation efficiency within the 
Milky Way. In this section, we expose the basic principles of the magnetically controlled 
theory of star formation.  

\subsubsection{Magnetic support}
The effect of the magnetic field is not easy to visualize because unlike the 
thermal pressure, it is highly non isotropic. In particular, the magnetic 
forces, $ {\bf j} \times {\bf B}$, where ${\bf j}$ is the electric current, vanishes
along the field lines. 
An easy way to estimate the magnetic support, is to compute the ratio of 
the magnetic over gravitational energies. For simplicity let us consider 
again a spherical and uniform cloud of mass $M$, volume $V$, radius $R$, threaded 
by an uniform magnetic field of strength $B$. 
The  magnetic flux within the cloud, $\psi$ is equal  to 
$\psi= \pi R^2 B$. As long as  the magnetic field remains well coupled 
to the gas (see next section), the magnetic flux threading the cloud will remain constant 
along time. 
The ratio of magnetic over gravitational energies for uniform density cloud threaded by 
a uniform magnetic field, is:
\begin{eqnarray}
{ E_ {\rm mag} \over E_{\rm grav}} = {B^2 V \over 8 \pi} \times {2  R \over 5 G M^2}
\propto {B^2 R^4 \over M^2} \propto \left( {\psi \over M } \right)^2.
\label{emag}
\end{eqnarray}
Remarkably, the ratio of magnetic over gravitational energies is independent of
the cloud radius. This implies that if the cloud contracts or expands, the
relative importance of these two energies remains the same. This is unlike the 
thermal energy of an isothermal gas, which becomes smaller and smaller compared to the gravitational
energy as the cloud collapses (e.g. eq.~\ref{ratio_ener}). 
It is clear from eq.~(\ref{emag}), that there is a critical value of the magnetic intensity for which 
the gravitational collapse is impeded even if the cloud was strongly compressed.  
Mouschovias \& Spitzer  (1976) have 
calculated accurately the critical value of the mass-to-flux ratio
 using the virial theorem and numerical calculations of the cloud bidimensional equilibrium.
A cloud which has a mass-to-flux ratio smaller (larger) than this critical value
cannot collapse and is called subcritical (supercritical).
It is usual to define $\mu = (M/\psi) / (M/\psi) _{\rm crit}$. Large values of 
$\mu$ correspond to small magnetic fields and thus supercritical clouds. 

Another important effect of the magnetic field is its ability to brake a rotating cloud. 
This is discussed rapidly later in this manuscript.

\subsubsection{Ambipolar diffusion}
At microscopic scales, the neutrals are not experiencing the 
Lorentz force which applies only on charged particles. Strictly speaking, this 
implies that at least two fluids  should be considered, the neutrals and the ions (in different contexts
 more than one fluid of charged particles must be considered),  
to treat the problem properly. Since the two fluids are coupled to each other by  collisions, 
the neutrals are nevertheless influenced by the magnetic field if the gas is sufficiently 
ionized. Since the ionization in molecular clouds is usually of the order 
of $10^{-7}$ or less, the density of the ions is much smaller than the density of the neutrals. 
It is  thus possible to neglect the inertia of the ions and assume mechanical equilibrium
between the Lorentz force and the drag force. This leads to:
\begin{equation}
{(\nabla \times {\bf B}) \times {\bf B} \over 4 \pi} = 
\gamma \rho \rho_i ( {\bf V}_i - {\bf V}),
\label{coupling}
\end{equation}
where $\rho_i$ and $\bf{V}_i$ are the ion density and velocity respectively, 
$\gamma \simeq 3.5 \times 10^{13}$ cm$^3$ g$^{-1}$ s$^{-1}$  is the drag coefficient
(Mouschovias \& Paleologou 1981). 
From eq.~(\ref{coupling}), the ion velocity can easily be expressed as a function 
of the neutral velocity and the Lorentz force. Considering now the induction equation, which entails
the velocity of the ions, and using eq.~(\ref{coupling}), we obtain
\begin{equation}
\partial _t {\bf B}  + \nabla \times  ({\bf B} \times {\bf V}) = 
\nabla \times \left( {1 \over 4 \pi \gamma \rho \rho_i}  ( (\nabla \times {\bf B}) \times {\bf B}) \times {\bf B} 
 \right). 
\label{induc}
\end{equation}
The left part of this equation is identical to the induction equation except that the  
 velocity of the  neutrals appears instead of the velocity of the ions. The right term 
 is directly responsible for the slip between the neutrals and the magnetic field. Although 
it is of the second order, it is not rigorously speaking a diffusion term. 
From this equation, it can easily be inferred a typical timescale for the ambipolar 
diffusion
\begin{eqnarray}
\tau_{\rm ad} \simeq {4 \pi \gamma \rho \rho_i L ^2 \over B^2 },  
\label{time}
\end{eqnarray}
where $L$ is the typical spatial scale relevant for the problem. In the context of 
star formation, $L$ could be the size of the cores, $R$. Ionization equilibrium 
allows to estimate that the ions density is about $\rho_i = C \sqrt{\rho}$, where 
$C=3 \times 10^{-16}$ cm$^{-3/2}$ g$^{1/2}$. 

If a dense core is initially subcritical (therefore magnetically supported), the diffusion 
of the field will progressively reduce the magnetic flux within the cloud. So after a
few ambipolar diffusion times, the cloud is becoming supercritical and the magnetic field 
is not able to prevent the collapse any more. 
The important and interesting question at this stage is,  how much is the collapse  
 delayed by this process? 
In order to estimate this time, it is usually  assumed that the cloud is in virial 
equilibrium that is to say: $B ^2 / 4 \pi   \simeq M \rho G / R$ (within a factor of a few) 
and to compute 
the ratio of $\tau_{\rm ad}$ and $\tau_{\rm ff}$, the freefall time (Shu et al. 1987).
This leads to:
\begin{eqnarray}
{ \tau _{\rm ad} \over \tau_{\rm ff}} \simeq 8.
\label{time_ratio}
\end{eqnarray}
Remarkably enough, $\tau _{\rm ad} / \tau_{\rm ff}$ is independent of $R$ and $M$, 
the size and mass of the cloud (as long as the virial assumption is verified). 
The important point is of course that $\tau _{\rm ad}$ is 
roughly 10 times higher than $\tau_{\rm ff}$. This implies therefore that 
ambipolar diffusion could possibly reduce the star formation rate significantly making it closer 
to the observed value.

\subsubsection{Predictions of ambipolar diffusion theory}
In order to make quantitative predictions, 
numerical simulations of magnetized collapse controlled by ambipolar diffusion,
 have been performed (e.g. Basu \& Mouschovias 1995).
These simulations are generally one dimensional and  assume a thin disk geometry.
They explore a wide range of magnetic intensities, from nearly 
critical cores, $\mu \simeq 1$, to very subcritical cores, $\mu \ll 1$. They also 
investigate the effect of changing the ionization which results in a weaker 
coupling between the magnetic field and the neutral gas. 
Velocity and density profiles potentially useful for comparison with observations, 
are therefore available in the literature. Here we simply draw some of the most 
important features. 

When the dense core is very subcritical, with values of $\mu$ as low as  $0.1$,
Basu \& Mouschovias  (1995)
find that the infall velocity in the outer part of the envelope is only a 
small fraction of the sound speed with values as low as $0.2 \times C_s$ whereas 
in the inner part, it gradually increases and reaches values of about 
$0.5-0.8 \times C_s$. The collapse is significantly delayed and occurs 
in about 15 freefall times. For nearly critical cores, $\mu \simeq 1$, the infall velocity 
is roughly twice higher than in  the previous case whereas the collapse occurs after 
about $\simeq$3 freefall times. Another interesting prediction, is the evolution of the central 
mass-to-flux ratio. The value of $\mu$ in the centre, smaller than 1 initially, grows with time
as the cloud loses its flux and eventually becomes larger than 1. By the time of  
protostar formation, typical values of $\mu$ are about $\simeq$2. Interestingly, this 
does not depend too much on the initial value of $\mu$. An important prediction of the
magnetically controlled models, is therefore that typically, values of $\mu$ around 2 
should be measured. Values significantly higher than 2, would certainly indicate 
that the collapse is not magnetically controlled.

Recently, Kunz \& Mouschovias (2009) proposed a theory of the 
initial mass function based on the development of gravitational 
instability within magnetically supported molecular clouds.

\subsection{Role of turbulence}
During the last two decades, the theory of magnetically controlled star formation, 
has  been challenged by 
a new theory based on supersonic turbulence (e.g. Mac Low \& Klessen 2004, 
McKee \& Ostriker 2007).
 The general idea of this  theory  is  that turbulence prevents most of the gas 
to collapse in a freefall time  and regulates the star formation, though different
aspects have been emphasized by various authors.

\subsubsection{Turbulent support and decay of turbulence}
Unlike magnetic field, it is not straightforward to anticipate the influence of
turbulence on the star formation rate. This is because, on one hand, the turbulent 
motions tend to spread out the  gas, reducing its ability to collapse, but on the other hand, 
the  turbulent motions may also  increase  the gas density locally when the flow is globally 
convergent. Moreover, the difficulty with theories involving turbulence is that it appears hopeless
to search for exact analytical descriptions, even for  highly idealized situations.
Therefore most of our theoretical knowledge of the interstellar turbulence is provided by
numerical simulations (e.g. Kritsuk et al. 2007, Schmidt et al. 2009). 
It appears nevertheless highly wishable to draw simple trends. 
To this purpose let us assume that the turbulence is sufficiently isotropic and 
is an additional support that can be 
described by a sound speed. Let $V _{\rm rms}$ be the root mean square of the velocity.
The effective sound speed of the flow is 
$C_{\rm s,eff} \simeq \sqrt{C_s^2 + V_{\rm rms}^2/3}$. Since the turbulence observed in 
molecular clouds is highly supersonic, $V_{\rm rms}^2$ is larger than $C_s^2$ by typically 
a factor of 25 to 100.
Therefore, $C_{\rm s,eff} \simeq  V_{\rm rms}/\sqrt{3}$
and the turbulent Jeans mass, $M_J  \propto V_{\rm rms}^3 / \sqrt{\rho}$.
On the other hand, turbulence creates density enhancements that can be estimated 
by the Rankine-Hugoniot jump conditions for an isothermal gas, $\rho / \rho_0 = (V/C_s)^2$, 
where $\rho_0$ is the mean cloud density.
Combining these two relations, we get the turbulent Jeans mass 
$M_J \propto V_{\rm rms}$, which  indicates that 
turbulence is globally supporting the cloud. This is certainly the case when turbulence 
is large and dominates over gravity. However the dual role of turbulence, which is also
compressing the gas through converging motions, requires a more complex treatment 
as presented in section~3.2.

To go further than these very simple analytic estimates, it is necessary to perform 
numerical calculations. Before  describing  some  results of these simulations, 
it is important to emphasize the fast decay of  turbulence which constitutes 
a severe issue of the turbulent theory. Consider a turbulent piece of fluid of 
size $L$, a robust conclusion seems to be that a significant fraction (say more than half)
of the initial turbulent energy is dissipated in about one crossing time, 
$L / V_ {\rm rms}$. This result, well established in the case of nearly incompressible fluids, 
has also been  inferred for numerical simulations of supersonic turbulence with and without 
magnetic field (Mac Low \& Klessen 2004).
Therefore, if not driven (that is to say when no external forcing is applied continuously to the flow),
 the turbulence decays quickly and 
thus cannot delay very significantly the collapse of a self-gravitating cloud.
In order to explain the low star formation rate in the Galaxy, the turbulence theory 
must invoke a driving source which continuously replenishes the turbulent energy.
Various sources of energies able to compensate for the dissipation have been 
proposed. A first category of models is based on feedback from stars due to their 
wind, jet or HII regions (e.g. Matzner 2002, 2007) 
while  a second category invokes the role 
of continuous accretion of diffuse atomic gas onto the molecular clouds
(Klessen \& Hennebelle 2010, Goldbaum et al. 2011).

Various numerical simulations have been performed to study directly the influence of 
turbulence on star formation. The influence of turbulence on the IMF is discussed
in the following sections (3 and 5), while below we restrict the discussion to its 
influence on the star formation rate. As it is not the main subject of these lectures, 
only a short and qualitative description is given.

\subsubsection{Hydrodynamical  turbulence}
As anticipated above, decaying turbulence cannot delay star formation significantly. 
Calculations done by Klessen et al. (2000) and Bate et al. (2003), indeed, show
that within a few freefall times, most of the gas has been accreted.

Driven turbulence can reduce the star formation rate if the driving is sufficient.
With a large scale driving (that is to say $k=1-2$ where $k$ counts
the number of driving wavelengths in the box), 
 providing an effective Jeans mass of 0.6 in a box which contains
 a total mass equal to 1 (that is to say that the box contains more than one turbulent Jeans mass), 
Klessen et al. (2000) find that more than half of the mass is accreted within 
one freefall time. The problem is 
less severe if the driving is on smaller scales ($k=3-4$ or $k=7-8$)
since in that case 3 to 6 (depending on the 
scales at which driving is applied) freefall times are needed to accrete half of the mass. 
If the driving is stronger, providing an effective Jeans mass of 3.2, these numbers 
are typically multiplied by a factor of 3 to 4. Star formation can be entirely suppressed
if sufficiently strong driving is applied at scales smaller than the thermal Jeans length inside
the box. Note that analytical estimates of the star formation rate have been derived by 
Krumholz \& McKee (2005), Padoan \& Nordlund (2011) and Hennebelle \& Chabrier (2011). 
While the first conclude that purely hydrodynamical turbulence can explain the very low 
efficiency of star formation within galaxies, the two others found much higher values, 
which are nevertheless in good agreement with star formation rate observed in 
molecular clouds (e.g. Heiderman et al. 2010) although possibly higher by a factor of a few.

\subsubsection{Turbulence in  magnetized clouds}
Supercritical magnetic fields are probably reducing the 
star formation rate by a factor of a few, that is to say the 
star formation rate is smaller in a cloud that has a 
supercritical magnetic field than in the otherwise identical 
non magnetized cloud (Price \& Bate 2009, Padoan \& Nordlund 2011)
although the exact physical mechanism by which this happens is not well understood.

Turbulence in subcritical clouds in the presence of ambipolar diffusion, 
has been investigated by Basu \& Ciolek (2004) and Li \& Nakamura (2005).  
These simulations combine the magnetic support 
and the turbulent motions. Since the clouds are initially subcritical, 
ambipolar diffusion plays an important role since it allows to reduce locally the 
magnetic flux. Interestingly, it has been found that in this context, turbulence
tends to accelerate the star formation. This is because, turbulence creates stiff gradients, 
due to the formation of shocks, in which the ambipolar diffusion takes place 
quickly. Indeed, eq.~(\ref{time}) shows that the ambipolar diffusion time decreases with the spatial 
scale, $L$. A very interesting result is that these simulations are able to reproduce the low star formation efficiency 
observed in the Milky Way provided the initial value of $\mu$ is small enough (typically $\mu \simeq 1$)
and the turbulence is sufficiently strong (typically $V_{\rm rms} \simeq 10 \times C_s$).

\section{Origin of the Initial mass function}

Several theories have been proposed to explain the origin of the IMF, invoking 
various physical processes that we tentatively classify in 
four categories: theories based on recursive fragmentation or pure 
gravity (e.g. Larson 1973), theories based on pure statistical argument, invoking 
the central limit theorem (e.g. Zinnecker 1984, Elmegreen 1997,  Adams \& Fatuzzo 1996), theories based on accretion
and, finally, theories invoking the initial Jeans mass in a fluctuating environment.  We will focus on the two latter
ones, which appear to be favored in the modern context of star formation.

\subsection{Theories based on accretion}

\subsubsection{Competitive accretion}

The theory of competitive accretion has been originally proposed
by Zinnecker (1982) and Bonnell et al. (2001).
It has then been used to interpret the series of numerical simulations 
similar to the ones performed by Bate et al. (2003).

The underlying main idea is that the accretion onto the stars is directly linked
to its mass in such a way that massive stars tend to accrete more efficiently
and thus become  disproportionally more massive than the low-mass stars.
The accretion rate is written as:
\begin{eqnarray} 
\dot{M}_* = \pi \rho V_{rel} R_{acc}^2,
\label{accret}
\end{eqnarray}  
where $\rho$ is the gas density, $V_{rel}$ is the relative velocity between the star and the gas
 while  $R_{acc}$ is the accretion radius. 
Bonnell et al. (2001) consider two situations, namely the cases where the gravitational
potential is dominated by the gas or by the stars. \\ \\

$\bullet$ Gas dominated potential \\
Let  $R$ be the spherical radius, $\rho$ the gas density and $n_*$ the number 
density of stars. 
In the gas dominated potential case, Bonnell et al. (2001) assume, following 
Shu (1977), 
that the gas density profile is proportional to $R ^{-2}$. They further 
assume that $n_* \propto R^{-2}$. The accretion radius, $R_{acc}$ is 
assumed to be equal to the tidal radius given 
by 
\begin{eqnarray} 
R_{tidal} \simeq 0.5 \left( {M_*\over M_{enc}} \right)^{1/3}  R,
\label{Rtidal}
\end{eqnarray}  
where $M_{enc}$ is the mass enclosed within radius $R$.
This choice is motivated by the fact that a fluid particle located at a distance
from a star smaller than $R_{tidal}$  is more sensitive to the star than to the cluster potential and will thus be accreted onto this 
star.

The mass of gas within a radius $R$, $M(R)$, is proportional to $R$ (since $\rho \propto
 R^{-2}$). The infall speed is about 
 $V_{in} \simeq \sqrt{G M(R) /R}$, and, assuming that the stars are virialized,
 one gets $V_{rel} \simeq V_{in}$.
The number of stars, $dN_*$, located between  $R$ and $R+dR$ is given by the relation 
$dN_* = n_*(R) \times 4 \pi R^2 dR \propto dR$. Thus 
eq.~(\ref{accret}) combined with the expression of 
$V_{rel}$ and $R_{acc}$ leads to the relation $\dot{M}_* \propto (M_*/R)^{2/3}$ and, 
after integration, 
 $M_* \propto R^{-2}$, $R \propto M_*^{-1/2}$. Consequently, we obtain:
\begin{eqnarray} 
dN \propto M_*^{-3/2} dM_*.
\label{dNacc1}
\end{eqnarray}  
Even though the mass spectrum is too shallow compared to the fiducial IMF, $dN/dM \propto M^{-2.3}$, it is interesting to see that this power law 
behaviour can be obtained from such a simple model. Note, however, that 
the model implies that stars of a given mass are all located in the same 
radius, which points to a difficulty of the model. \\ \\

$\bullet$ Star dominated potential \\
When the potential is dominated by the stars located at the centre of the cloud,  
the density is given by $\rho \propto R^{-3/2}$, which corresponds to the expected density 
distribution after the rarefaction wave has propagated away
(Shu 1977). The velocity is still assumed to be $V_{rel} \propto R^{-1/2}$. 
The accretion radius is now supposed to correspond to the Bondi-Hoyle radius 
as the gas and the star velocities are no longer correlated.
This leads to
$\dot{M}_* = \pi \rho V_{rel} R_{BH}^2$, where $R_{BH} \propto M_* / V_{rel}^2$. 
It follows
\begin{eqnarray} 
\dot{M}_* \propto M_*^{-2}.
\label{dMacc2}
\end{eqnarray} 
One can then show (Zinnecker 1982, Bonnell et al. 2001) that under reasonable assumptions,
 $dN \propto M_*^{-2} dM_*$. This estimate is in better agreement with the 
Salpeter exponent, although still slightly too shallow. These trends 
seem to be confirmed by the simulations performed by Bonnell et al. (2001)
which consists in distributing 100 sink particles in a cloud of total mass
about 10 times the total mass of the sinks initially (their fig. 3). \\ \\

$\bullet$ Difficulties of the competitive accretion scenario \\
As obvious from the previous analytical derivations, finding an explanation for 
the Salpeter exponent with the competitive accretion scenario appears to be difficult, 
even though numerical simulations (as the ones presented in Bonnell et al. 2001) seem to successfully achieve this task.
However, although the IMF exponent 
is close to the Salpeter one in the star dominated potential case,  this scenario entails by construction the Bondi-Hoyle accretion, which  is at least a factor ~3 lower than the mass infall rate resulting from gravitational collapse at the class-0 and I stages and leads to too long accretion times compared with observations (Andr\'e et al. 2007, 2009). Another difficulty
of this scenario is that it does not explain the peak of the IMF which 
might be related to the Jeans mass (see \S~\ref{jeans}). Finally, 
it is not clear that competitive accretion can work 
in the case of non-clustered star formation, for which the gas density
is much too small. As no evidence for substantial IMF variation  among different regions
has yet been  reported (e.g. Bastian et al. 2010), this constitutes a  difficulty for this model as a general model for star formation.
Perhaps this scenario applies well to the formation of massive stars.

\subsubsection{Stopped accretion}
The principle of this type of models is to assume that the accretion of gas onto 
the stars or the dense cores is a non-steady process, which stops because of either the finite reservoir of mass
or the influence of an outflow which sweeps up the 
remaining gas within the vicinity of the accreting protostar.

The first studies were performed by Silk (1995) and Adams \& Fatuzzo (1996). 
They first relate the mass of the stars to the physical parameters of the cloud such as
sound speed and rotation and then assume that an outflow whose properties 
are related to the accretion luminosity stops the cloud collapse. 
Using the Larson (1981) relations, they can link all these parameters to 
the clump masses. Since the mass spectrum of these latter is known (e.g. Heithausen et al. 1998),
they infer the IMF.

A  statistical approach has been carried out by Basu \& Jones (2004).
These authors assume that the dense core distribution is initially 
lognormal, justifying it by the large number of processes that control their 
formation (and invoking the central limit theorem). Then, they argue that 
the cores grow by accretion and postulate that the accretion rate is 
simply proportional to their mass, 
$\dot{M} = \gamma M \rightarrow M(t)=M_0 \exp(\gamma t)$,
leading  to $\log M = \mu = \mu_0 + \gamma t$. Finally, they assume that accretion 
is lasting over a finite period of time given by 
$f(t)=\delta \exp(-\delta t)$.
The star mass distribution is thus obtained by summing over the 
accretion time distribution.
\begin{eqnarray}
f(M) &=& \int _0 ^\infty { \delta \exp(-\delta t) \over \sqrt{2 \pi} \sigma_0 M } \exp \left( -{ (\ln M - \mu_0 -\gamma t)^2\over 2 \sigma_0^2}  \right) dt \\
&=& {\alpha \over 2} \exp( \alpha \mu_0 + \alpha^2 \sigma_0^2/2) M^{-1-\alpha} {\rm erf} \left( {1 \over \sqrt{2} } (\alpha \sigma_0 - { \ln M - \mu_0 \over \sigma_0})\right),
\nonumber
\end{eqnarray}
where $\alpha=\delta/\gamma$ and $\sigma_0$ characterizes the width of the initial dense
core distribution. As  $\delta$ and $\alpha$ are controlled by the same types 
of processes, their ratio is expected to be of the order of  unity
and thus $f(M)$ exhibits a powerlaw  behaviour close to the fiducial IMF. 

In a recent study, Myers (2009) develops similar ideas in  more details, 
taking  into account the accretion coming from the surrounding 
background. Adjusting two parameters, he reproduces quite nicely the observed 
IMF (his figure 5).

A related model has also been developed by Bate \& Bonnell (2005) based on an 
idea proposed by Price \& Podsiadlowski (1995). They consider objects
that form by fragmentation within a small cluster and are ejected 
by gravitational interaction with the other fragments, which stops the accretion process. 
Assuming a lognormal accretion rate and an exponential probability of being ejected, 
these authors construct a mass distribution that can fit the IMF for some choices of parameters. 

In summary, the stopped accretion scenario presents interesting ideas and, providing (typically 2 or more) adequate adjustable parameters, 
can reproduce reasonably well the IMF. However, the very presence of such parameters, which characterizes our inability to precisely determine the processes that halt accretion,
illustrates the obvious difficulties of this class of models.

\subsection{Gravo-turbulent theories}
\label{jeans}

While  turbulence is not determinant in the accretion models, 
it is one of the essential physical processes for the two theories presented in this 
section, although the  role it plays differs in both models, as shown below. 
The theories proposed along this line
seemingly identify cores or {\it pre-cores} and are motivated 
by the strong similarity between the observed core
 mass function (hereafter) CMF and the IMF (e.g. Andr\'e et al. 
2010).

The first theory that combined turbulence and gravity was
proposed by Padoan et al. (1997). In this paper, the authors consider
a lognormal density distribution - density PDF computed from
 numerical simulations (e.g. V\'azquez-Semadeni 1994, Kritsuk et al. 2007, Schmidt et al. 2009, Federrath et al. 2010a) 
are indeed nearly lognormal - and select the regions of the flow which 
are Jeans unstable.
By doing so, they get too stiff an IMF (typically 
$dN/ dM \propto M^{-3}$) but nevertheless find a lognormal behaviour at small 
masses, a direct consequence of the lognormal density distribution,
 and a powerlaw one at large masses.

\subsubsection{Formation of cores by MHD shocks}
The idea developed by Padoan \& Nordlund (2002) is slightly different. 
These authors consider a compressed layer formed by ram pressure in a weakly 
magnetized medium. They assume that the magnetic field is parallel to 
the layer and thus perpendicular to the incoming velocity field. 
The postshock density, $\rho_1$, the thickness of the layer, $\lambda$,
and the postshock magnetic field, $B_1$, can be related to the 
Alfv\'enic Mach numbers, ${\cal M}_a=v/v_a$ 
($v$ is the velocity and $v_a$ the Alfv\'en speed),
  and preshocked quantities, $\rho_0$ and $B_0$ according to the shock conditions:
\begin{eqnarray}
\rho_1/\rho_0 \simeq {\cal M}_a \; , \; \lambda / L \simeq {\cal M}_a^{-1} \; , \; B_1/B_0 \simeq {\cal M}_a \; ,
\label{eq_pn1}
\end{eqnarray}
where $L$ is the scale of the turbulent fluctuation. Note that for classical 
hydrodynamical isothermal shocks, the jump condition is typically $\propto {\cal M}^2$. 
The dependence on ${\cal M}_a$ instead of ${\cal M}^2$ stems from the 
magnetic pressure which is quadratic in $B$. As we will see, this 
is a central assumption of this model. 

The typical mass of this perturbation is expected to be
\begin{eqnarray}
M \simeq \rho_1 \lambda^3 \simeq \rho_0 {\cal M}_a \left( {L \over {\cal M}_a } \right)^3 \simeq \rho_0 L^3  {\cal M}_a^{-2}.
\label{eq_pn2}
\end{eqnarray}
As the flow is turbulent, the velocity distribution depends on the scale 
and $v \simeq L^\alpha$, with $\alpha=(n-3)/2$,
 $E(k) \propto k^{-n}$ being the velocity powerspectrum \footnote{$n$ is denoted $\beta$ in Padoan \& Nordlund (2002), more precisely $n-2=\beta$}.
Combining these expressions with eq.~(\ref{eq_pn2}), they infer
\begin{eqnarray}
M \simeq {\rho_0 L_0^3 \over {\cal M}_{a,0} } \left( { L \over L_0} \right)^{6-n}, 
\label{eq_pn3}
\end{eqnarray}
where $L_0$ is the largest or integral scale of the system and ${\cal M}_{a,0}$
the corresponding Mach number. To get a mass spectrum, it is further assumed 
that the number of cores, $N(L)$, formed by a velocity fluctuation of scale $L$, 
is proportional to $L^{-3}$. Combining this last relation with eq.~(\ref{eq_pn3})
leads to
\begin{eqnarray} 
{dN \over d \log M}  \simeq M^{-3/(6-n)}.
\label{eq_pn4}
\end{eqnarray}
For a value of $n=3.74$ (close to what is inferred from 3D numerical simulations), 
one gets $dN/d \log M \simeq M^{-1.33}$, very close to the Salpeter exponent.

So far, gravity has not been playing any role in this derivation and the 
mass spectrum that is inferred is valid for arbitrarily small masses.
In a second step, these authors consider a distribution 
of Jeans masses within the clumps induced by turbulence. As the density 
in turbulent flows presents a lognormal distribution, they {\it assume} 
that this implies a lognormal distribution of Jeans lengths and they 
multiply the mass spectrum (\ref{eq_pn4}) by a 
distribution of Jeans masses, which leads to
\begin{eqnarray} 
{dN \over d \log M} \simeq M^{-3/(6-n)} \left( \int _0^M p(m_J) dm_J \right).
\label{eq_pn5}
\end{eqnarray}
The shape of the mass spectrum stated by eq.~(\ref{eq_pn5}) is very similar to 
the observed IMF (see for example the figure 1 of Padoan \& Nordlund 2002).

Note, however, that  difficulties with this theory have been pointed out by 
McKee \& Ostriker (2007) and Hennebelle \& Chabrier (2008).
Eqn~(\ref{eq_pn1}), in particular, implies that in the densest regions
where dense cores form, the magnetic field 
is proportional to the density, in strong contrast with what is observed both in simulations
(Padoan \& Nordlund 1999, Hennebelle et al. 2008) and in
observations (e.g. Troland \& Heiles 1986). This is a consequence of 
the assumption that the magnetic field and the velocity field are perpendicular,
which again is not the trend observed in numerical simulations.
In both cases, it is found 
that at densities lower than about 10$^3$ cm$^{-3}$, $B$ depends only weakly 
on $\rho$ while at higher densities, $B \propto \sqrt{\rho}$. This constitutes a  
problem for this theory, as the index of the power law slope is a direct consequence of 
eq.~(\ref{eq_pn1}). Assuming a different relation between $B$ and $\rho$,  as the aforementioned observed one, would 
lead to a slope stiffer than the Salpeter value. Furthermore, the Salpeter IMF is recovered
in various purely hydrodynamical 
simulations (e.g. Bate et al. 2003), while the Padoan \& Nordlund theory 
predicts a stiffer distribution ($dN/ dM \propto M^{-3}$) in the hydrodynamical case.

\subsubsection{Turbulent dispersion}
Recently, Hennebelle \& Chabrier (2008, 2009) proposed 
a different theory which consists in counting the mass of the fluid regions
within which gravity dominates over the sum of all supports, thermal, turbulent, and 
magnetic, according to the virial condition (see also Hopkins 2012). 
In this approach, the role of turbulence
is dual: on one hand it promotes star formation by locally compressing the gas but on the 
other hand, it also quenches star formation because of the turbulent 
dispersion of the flow, which is taken into account in the selection of the pieces of fluid that collapse. 

Here instead of presenting the full formalism that can be found in 
Hennebelle \& Chabrier (2008), we qualitatively describe 
the procedure and important 
physical features and we then focus on simpler scaling relations
that show how the Salpeter slope can be simply recovered using the 
turbulent support (Chabrier \& Hennebelle 2011).
The  theory is formulated by deriving an extension of the Press \& Schechter (1974) 
statistical formalism, developed in cosmology.
The principle of the method is the following. First, the density field is 
smoothed at a scale $R$, using a window function. Then, the total mass 
contained in  areas which, at scale $R$,  have a density contrast larger
 than the  specified density  criterion $\delta_R^c$,  is obtained by
 integrating accordingly the density PDF. 
This mass, on the other hand, is also equal to the total mass located in
 structures
of mass larger than a scale dependent critical mass $M_R^c$, which will end 
up forming structures of mass {\it smaller than or equal} to this critical
 mass for collapse. 
The two major  differences are (i) the underlying density field, characterized by
small and Gaussian fluctuations in the cosmological case while  lognormal in the 
star formation case, and (ii) the selection criterion, a simple scale-free density 
threshold in cosmology while scale-dependent, based on the virial theorem in the second case.
That is,  fluid particles which satisfy the criterion
\begin{eqnarray}
  \langle V_{\rm rms}^2\rangle    + 3\, (C_s^{eff})^2 < - E_{\rm pot} / M
\label{viriel_ceff}
\end{eqnarray}
are assumed to collapse and form a prestellar bound core.
The turbulent rms velocity obeys a power-law correlation with the size of the
region, the observed so-called Larson relation,
with $V_0\simeq 1\, {\rm km\, s}^{-1}$ and $\eta \simeq 0.4$-0.5 (Larson 1981).

The theory is controlled by two Mach numbers, namely
a Mach number (called ${\cal M}_*$) 
defined as the non-thermal velocity to sound speed ratio at the mean   Jeans 
scale $\lambda_J^0$ (and not at the local Jeans length), and the usual Mach
 number, ${\cal M}$,
which represents the same quantity at the scale of the turbulence injection
 scale, $L_i$, assumed to be the characteristic size of the system,
The global Mach number, ${\cal M}$,  broadens the density PDF, as 
$\sigma^2 = \ln (1 + b^2 {\cal M}^2)$,
illustrating the trend of supersonic turbulence to promote star formation by 
creating new overdense collapsing seeds. 
 The effect described by ${\cal M}_*$  is the  additional non thermal support
induced by the turbulent dispersion.
In particular, at large scales the net effect of turbulence is to stabilize
 pieces of fluid that would be gravitationally unstable if only the thermal 
support was considered.

The condition to select a  marginally unstable 
piece of fluid of mass $M$ and size $R$,  with internal velocity 
dispersion $\sigma_R \equiv \sigma$, simply stems from the virial 
condition: $2E_{kin}+E_{pot}=0$, 
where $E_{kin}\sim (1/2)\rho \sigma^2 $ denotes
the kinetic energy and $E_{pot}\sim \rho G\frac{M}{R}$ the gravitational 
energy. Using dimensional analysis, this yields the relation
\begin{eqnarray}
 M\sigma^2\sim G \frac{M^2}{R}{\hskip .5cm}{\rm i.e.} {\hskip .5cm}M\sim 
{R\sigma^2 \over G}\,.
\label{vir}
\end{eqnarray}

\noindent In the limit where the velocity dispersion is dominated by thermal motions, and assuming nearly isothermal conditions, $\sigma$ does not depend on $R$ and condition (\ref{vir}) yields
\begin{eqnarray}
 M_{th}\propto R{\hskip 2.8cm}({\rm THERMAL\,\,SUPPORT}).
\label{therm}
\end{eqnarray}

\noindent Conversely, in the limit where the velocity dispersion is dominated by turbulent motions,  $\sigma \propto R^{{(n-3)\over 2}}$, yielding for the marginal instability condition

\begin{eqnarray}
M_{turb} \propto  R^{(n-2)} \propto R^{2\eta+1}{\hskip .2cm}({\rm TURBULENT\,\,SUPPORT}).
 \label{Mturb}
\end{eqnarray}
with $\eta=(n-3)/2$. 
The thermal case is recovered for $n=3$, i.e. $\eta=0$. For $n>3$, i.e. $\eta>0$, we note the stronger dependence of the mass $M$ upon the size $R$ in the presence of turbulent support, compared with the
pure thermal case (eq.~\ref{therm}). The complete relation is obtained in Hennebelle \& Chabrier (2008; their eq.(26)) and illustrates that a bound stable region can achieve larger masses for a given size
thanks to the non-thermal support at the characteristic Jeans length, naturally leading to a larger number of high mass reservoirs (then of massive prestellar cores), which would otherwise have collapsed long before. This impact of nonthermal support on the mass-size relation of massive bound cores has been confirmed by recent
numerical simulations (Schmidt et al. 2010, their fig.7).

 Assuming a large enough multi-scale structured space of size $L$ and  dimension $D$, thus volume $L^D$, 
 the probability of having $N$ gravitationally bound structures on a scale larger than $R$ 
is given by the probability law ${\cal P}(N_{R})\equiv N(>R)\propto R^{-D}$.
In Fourier space, this simply means that the number of fluctuations of wavenumber $k\sim 1/R$ is proportional to the volume of a fluctuation, i.e. $dN(k)\propto d{\vec k}=k^{D-1}dk$.
The density probability, i.e. the probability to have a structure of size $\in[R,R+dR]$, thus reads as
\begin{eqnarray}
\frac{dN(R)}{dR}\propto R^{-(1+D)}. \label{N}
\end{eqnarray}
A homogeneous volume in 3 dimensions obviously corresponds to $D=3$, i.e. a uniform density probability $dN(k)/d{\vec k}=constant$, as assumed e.g. in Padoan \& Nordlund (2002).

From Eqs.~(\ref{Mturb}) and (\ref{N}), the probability of having a bound structure of mass $\in[M(R),M+dM(R+dR)]$, which defines the CMF/IMF, reads
\begin{eqnarray}
\frac{dN}{dM}&\propto& M^{-\alpha}\nonumber \\
{\rm with}\,\,\,\alpha&=&{n+1 +(D-3) \over n-2} ={2\eta+4 +(D-3) \over 2\eta+1} 
\label{Nturb}
\end{eqnarray}

As seen from Eq.(\ref{Nturb}), for $D=3$, pure thermal support ($n=3,\eta =0$) yields a power-law IMF $\frac{dN}{dM}\propto M^{-4}$, substantially steeper than the 
Salpeter value, $\alpha=2.35$\footnote{Assuming a Gaussian (lognormal) density probability instead of a uniform one yields in that case $\frac{dN}{dM}\propto M^{-3}$ (eq.~(37) of HC08).}.
In the two limiting cases of incompressible (Kolmogorov) ($n=11/3,\eta =1/3$) and pressureless (Burgers) ($n=4,\eta =1/2$) turbulence, the turbulent support yields
$\alpha=2.8$ and $\alpha=2.5$, respectively. In molecular clouds, turbulence is assumed to scale according to the observed Larson (1981) velocity dispersion - size relation
with a typical value $\eta\sim 0.4$-0.5. This value is well recovered for the aforementioned power spectrum index inferred from numerical simulations of supersonic turbulence, $n=3.8$.
According to eq.~(\ref{Nturb}), this yields a slope for the CMF/IMF $\alpha=2.66$,
between the Kolmogorov and Burgers values. Note that the index $D$ observed in molecular clouds
is typically smaller than $D=3$ and closer than $D\simeq 2.5$ which leads to values for $\alpha \simeq 2.2-2.5$
very close to the Salpeter exponent (see Chabrier \& Hennebelle 2011) for details.

Comparisons with the Chabrier (2003) IMF have been performed for a series of cloud parameters 
(density, size, velocity dispersion) and good agreement has been found 
(Hennebelle \& Chabrier 2009) for clouds typically 3 to 5 times denser than the 
mean density inferred from Larson (1981) density-size relation.
Comparisons 
with numerical simulations have also been performed. In particular, 
Schmidt et al. (2010), performing supersonic isothermal simulations with various 
forcing, have computed the mass spectrum of cores supported 
either by pure thermal support or by turbulent plus thermal support.
  Their converged simulations show very good {\it quantitative} agreement 
with the present theory, confirming that turbulent support is needed to 
yield the Salpeter index. Note that Schmidt et al. (2010) use for the density 
PDF the one they measure in their simulations which is nearly, but not exactly lognormal.
Comparisons with the results of SPH simulations 
(Jappsen et al. 2005) including self-gravity and thermal properties of the gas have also
been found to be quite successful (Hennebelle \& Chabrier 2009).

\subsubsection{Difficulties of the gravo-turbulent theories}
One natural question about any IMF theory is to which extent it varies with 
physical conditions. Indeed, there is strong observational support for a nearly invariant form and peak location of the IMF in various environments under Milky Way like conditions (see e.g. Bastian et al. 2010).
Jeans length based theories could
have difficulty with the universality of the peak position, since
 it is linked to the Jeans mass which varies with the gas density. Various propositions
have been made to alleviate this problem. Elmegreen et al. (2008) and Bate (2009) propose
that the gas temperature may indeed increase with density, resulting in a
Jeans mass which weakly depends on the density, while Hennebelle \& Chabrier (2008) propose that 
for clumps following Larson relations, there is a compensation between the 
density dependence of the Jeans mass and the Mach number dependence 
of the density PDF, resulting in a peak position that is insensitive to the clump
size.

A related problem is the fact that massive stars are often observed to be located in the 
densest regions, where the Jeans mass is smaller. Indeed, $M_J \propto \rho^{-1/2}$ when a purely thermal support is considered, whereas
$M_J \propto \rho^{-2}$ when turbulence is taken into account (assuming 
that $V \propto L^{0.5}$).
 This constitutes a difficulty for theories based on Jeans mass 
although, as seen above, the issue is much less severe when turbulent support is considered
as massive stars can be formed at densities only few times smaller than the densities
at which low-mass stars form. Another possibility is that dynamical interactions 
between young protostars may lead to the migration of massive stars in the 
center of the gravitational well.

Furthermore, the dependence of the freefall time on the Jeans mass
should also modify the link between the CMF and the IMF, as
pointed out by Clark et al. (2007). 
This is particularly true for theories which invoke only  thermal 
support. When  turbulent support is included,
 the free-fall time is found to depend only weakly on the mass, with
$t_{ff}\propto M^{1/4}$ (see McKee \& Tan 2003 and Hennebelle \& Chabrier 2009 App. C), 
resolving this collapsing time problem. 
We stress that this time represents 
the time needed for the whole turbulent Jeans mass to be accreted. 
It is certainly true that, within this turbulent Jeans mass, small 
structures induced by turbulent compression will form rapidly.

Generally speaking, the fragmentation that occurs during the collapse
could constitute a problem for theories invoking Jeans masses.
Although this problem is far 
from being settled, it should be stressed that 
while the presence of small 
self-gravitating condensations induced by turbulence 
 at the early stages of star formation and 
embedded into larger ones is  self-consistently taken into account 
in the turbulent dispersion theory, 
the fragmentation induced by gravity, i.e. by the dense regions produced
during collapse (see section~5.2.2) is not included in this theory. 
It certainly 
constitutes another problem if this mode turns out to be dominant.

\section{Formation of brown dwarfs through disk fragmentation}
In the previous section, the mechanism by which stars and  brown dwarfs formed, 
namely the selection by gravity of dense fluctuations induced by turbulence, 
was identical. Other scenarii have been proposed in which  the formation of brown dwarfs 
is different. This is in  particular the case for the fragmentation of 
massive disks as recently emphasized by  Whitworth \& Stamatellos (2006) 
and Matzner \& Levin (2005). It is the purpose of this section to present 
this model.

\subsection{Basics}
Here we start by presenting the basics of 
disk physics in particular the processes
which have been used by these authors to infer their theories.

\subsubsection{Rotational support}
Let us consider a cloud of mass, $M$, rotating at a rate $\Omega$.
If angular momentum is conserved, we have the relation $R^2 \Omega = j$, where $j$
is the specific angular momentum.
The rotation energy is proportional  to $ M R^2 \Omega^2$ and thus
\begin{eqnarray}
{E_{\rm rot} \over E_{\rm grav}} \propto { M R^2 \Omega^2 \over M^2 G / R} 
= { j^2 \over G  M} {1 \over R}. 
\label{rap_rot_grav}
\end{eqnarray}
Thus, as the cloud collapses, the radius $R \rightarrow 0$ and 
the energy ratio increases. This shows that the centrifugal support 
becomes dominant even if rotation is small initially compared 
to gravity. This well known and very important behaviour 
is called the centrifugal barrier. It is responsible for disk formation 
in astrophysics and makes the transport of angular momentum a crucial 
issue. Indeed, if angular momentum was conserved, eq.~(\ref{rap_rot_grav})
shows that stars would probably never form.

\subsubsection{Centrifugal radius and disk growth}
Let us consider a star of mass $M$ and a fluid particle whose 
specific angular momentum is $j$. The centrifugal radius, $r_d$, is 
simply obtained when centrifugal and gravitational forces on the fluid particle
are equal, which leads to
\begin{eqnarray}
r_d = {j^2 \over GM}.
\label{centrifug}
\end{eqnarray}
This expression shows that the disk formation in a collapsing cloud, crucially depends on 
the specific angular momentum distribution. To illustrate this more quantitatively
let us consider a cloud in solid body rotation, 
whose  density  profile is initially proportional to $1/R^2$ (e.g. Shu 1977).
In this case, the mass enclosed in a sphere of radius $R_0$ is $M(R_0) \propto R_0$
while the specific angular momentum $j(R_0) = R_0^2 \Omega$. 
Therefore $j(R_0) \propto M(R_0)^2$. As accretion proceeds, 
the  shells of increasing $R_0$ are reaching the central part, say the star-disk 
system. The centrifugal radius for a fluid particle
initially located in $R_0$ is given by $r_d = j(R_0)^2 / (G M(R_0)) \propto M(R_0)^3$
(Terebey et al. 1984).

On the other hand, for a cloud in solid body rotation for which 
 the density profile is initially flat, i.e. the density is uniform, 
we have $M(R_0) \propto R_0^3$ and $j \propto M(R_0)^{2/3}$ leading to 
$r_d \propto M(R_0)^{1/3}$. As the mass delivered within the cloud 
is typically of the order of a few $C_s^3/ (G) t$, we see that 
while in the first case, the disk grows like $t^{3}$, i.e. extremely slowly
(since $t \simeq 0$ initially), it grows much faster in the second case.

\subsubsection{Equations of thin disk  and steady accretion}
After averaging along the disk axis, the continuity and 
angular momentum conservation  equations, one obtains (see e.g. Pringle 1981)
\begin{eqnarray}
\partial_t \Sigma  + \partial _r(r V_r \Sigma) =0,
\label{continuity}
\end{eqnarray}

\begin{eqnarray}
  \partial_t (\Sigma r^2 \Omega)  +  {1 \over r} \partial _r  ( r^3 \Omega \Sigma V_r)  
= {1 \over r} \partial _r  ( \nu r^3 {d \Omega \over dr} \Sigma),
 \label{angular}
\end{eqnarray}
where $\Sigma$ is the column density through the disk, $V_r$ is the radial velocity, 
and $\nu$ is the viscosity. While the left hand terms in eq.~(\ref{angular}) 
are the standard conserved angular momentum and angular momentum flux, 
the right hand side is the torque exerted  through  viscosity by the 
neighboring annulus.

In steady state, eq.~(\ref{angular}) leads to $\Omega V_r = \nu d \Omega / dr$.
As the mass flux $\dot{M}= 2 \pi r \Sigma V_r$, we get
$\dot{M}= 2 \pi \nu r d \Omega / dr$. For a Keplerian disk, mechanical equilibria 
in radial and axial directions lead to $GM / r^2 \simeq r \Omega^2 $ and 
$G M h / r^3 \simeq c_s^2 / h$ and thus $h \simeq c_s/\Omega$. It is usual to introduce 
the adimentional quantity, $\alpha= \nu / (c_s h)$, which leads to the relation
\begin{eqnarray}
\dot{M} = 3 \pi \alpha {\Sigma c_s^2 \over \Omega}.
 \label{dM_dt}
\end{eqnarray}

\subsubsection{Thermal balance}
In the absence of other source, the heating is due to the viscous dissipation and is 
given by (e.g. Balbus \& Hawley 1998, Hartmann 2009)
\begin{eqnarray}
Q_+ = (\nu \Sigma r^2 d \Omega / dr) \times d \Omega / dr = \nu \Sigma \left( r {d \Omega \over dr } 
\right)^2
= \alpha \Sigma c_s^2 \Omega \left( { d \log \Omega \over d \log r}\right)^2,
\label{heating}
\end{eqnarray}
which is simply the rate of work per units of length of the viscous torque.
For a Keplerian disk this becomes
\begin{eqnarray}
Q_+ = {3 \over 4 \pi} \Omega^2 \dot{M}.
\label{heating_kep}
\end{eqnarray}
If $\tau_{cool}$ is the cooling time, the cooling rate is about
\begin{eqnarray}
Q_- = {U _{th} \over t_{cool} } = {\Sigma c_s^2 \over \gamma (\gamma-1) t_{cool}},
\label{cooling}
\end{eqnarray}

At thermal equilibrium heating and cooling compensate each other and thus
$Q_+ = Q_- $ leading to 
\begin{eqnarray}
\alpha =  \left( { d \log \Omega \over d \log r } \right)^{-2} 
{1 \over \gamma (\gamma-1) \Omega t_{cool}},
\label{balance_therm}
\end{eqnarray}
Thus, disks that cool fast must have a large  viscosity to provide enough 
heat to compensate the efficient radiative loss.

\subsubsection{The Toomre criterion}
The stability of a centrifugally supported disk is governed by 
the Toomre criterion $Q = C_s \kappa / (\pi G \Sigma)$ where 
$\kappa^2 = 4 \Omega^2 + R d\Omega^2 / dR$ is the epicyclic frequency and $\Sigma$ is the 
column density through the disk (Toomre 1964).
The complete derivation of this relation can be found in 
Binney \& Tremaine (1987) and is beyond the scope of the
present manuscript where we follow a more phenomenological 
but simpler approach. 

Let us consider a piece of fluid of size $\delta R$ and mass $\delta M=\pi \delta R^2 \Sigma$.
Its thermal, gravitational and rotational energies are about 
$\delta E_{th} \simeq c_s^2 \delta M  \propto \delta R^2$, 
$\delta E_{gr} \simeq G \delta M^2 / \delta R \propto \delta R^3 $ and 
$\delta E_{rot} \simeq \delta M (\Omega \delta R)^2 \propto \delta R^4$.
Thus it appears that while at small scales, when $R \le R_{th} = c_s^2 / (\pi G \Sigma )$,
 the disk is stabilized by thermal support, 
it is stabilized at large scales when $R \ge R_{rot}=  \pi G \Sigma  / \Omega^2 $ by the rotational one.
Thus only the intermediate scales between $R_{th}$ and $R_{rot}$ are possibly unstable.

The condition for disk stability is that $R_{th} > R_{rot}$ in which case all scales
are stabilized either by thermal or by rotational support. This leads to 
$ c_s^2 / (\pi G \Sigma ) \ge  \pi G \Sigma  / \Omega^2 $ and finally 
$Q =  c_s \Omega / (\pi G \Sigma) \ge 1$, which is almost, although not exactly,
 the famous Toomre's criterion as the epicyclic frequency replaces  the rotation,
except for a Keplerian disk for which $\kappa = \Omega$.

\subsubsection{Self-gravitating disks and self-regulation}
So far, it has been assumed that an effective viscosity 
was responsible of the transport of angular momentum
without attempting to relate it to a physical process. 
When the disk is self-gravitating the angular momentum 
conservation equation can be written as 
(e.g. Lodato \& Rice 2004)

\begin{eqnarray}
\partial_t(\Sigma r^2 \Omega)+ {1 \over r} \partial_r (r^3 \Sigma V_r \Omega +
T_{r \phi}) = 0
\label{grav_stress}
\end{eqnarray}
where $T_{r \phi} = \int {r^2 g_\theta g_r \over 4 \pi G} dz 
 + \Sigma \delta V_r \delta V_\phi$ and $\delta V_r = V_r - <V_r> $,
$\delta V_\phi = V_\phi - <V_\phi> $ also known as a Reynolds decomposition.  
The first is obtained by using the Poisson equation to write 
the gravitational term as $(\rho / r) \partial _\theta \phi = 
 (r^{-1} \partial_r ( r \partial_r \phi) + \partial_\theta^2 \phi / r^2 
 + \partial_z^2 \phi ) \partial_\theta \phi / (4 \pi G r)$
and integrating along the z-axis. The second one is the Reynolds stress 
and comes from the Reynolds decomposition of the velocity field.

Equation~(\ref{grav_stress}) suggests that the gravitational stress provides
an effective viscosity which is due to the gravitational field itself
but also to the non-axisymmetric velocity fluctuations that it
generates. One important difference however with a simple 
viscosity arises from the non-local nature of gravity that could in principle
deeply modify its behaviour. It seems however
that gravitational stress behaves in a way which closely 
resembles a local viscous stress (Lodato \& Rice 2004).
This has important implications because 
according to the Toomre criterion, when the disk is cold, it is unstable. 
Thus, non-axisymmetric spiral modes develop making the gravitational 
stress higher and thus according to eq.~(\ref{heating}) the heating increases
and thus the disk becomes more stable. This suggests that self-gravitating disks 
self-regulate (Paczynski 1978), that is they tend to maintain their Toomre parameter, $Q$
around 1. This has been extensively observed in numerical simulations
as for example by Lodato \& Rice (2004). Varying the initial parameters 
of their disks, they find that except close to the center or the edge, $Q$
stays typically between 1 and 1.5.

\subsubsection{Non-linear stability}
The analysis of the previous section shows that the disk stability 
is ultimately linked to its ability to cool. This 
has been studied in details by Gammie (2001) who run 
a series of disk simulations (using the shearing box approximation) 
in which he varies the cooling time. Indeed, eq.~(\ref{balance_therm})
shows that a short cooling time implies a large effective 
viscosity and thus a large gravitational stress. As the latter 
is the consequence of non-axisymmetric perturbations, it is intuitive 
that large values of the $\alpha$ parameter will tend to be associated
with very perturbed and therefore very unstable disks.
Gammie (2001) shows that the critical value for the cooling time 
is about $t_{cool} \simeq 3 \Omega^{-1}$. Below this value 
the disk fragments while it remains stable above.
Note that the exact value of the coefficient 
$t_{cool} \Omega$ remains 
controversial (e.g. Meru \& Bate 2011).

\subsection{Brown dwarf formation  through disk fragmentation}

\subsubsection{Basic scenario}
The formation of brown dwarfs through disk fragmentation has been 
recently investigated by Matzner \& Levin (2005) and Whitworth \& Stamatellos
 (2006), using similar although not identical arguments. Below we 
present  the arguments of Matzner \& Levin (2005) and mention the difference 
with the ones used by Whitworth \& Stamatellos (2006).

Matzner \&  Levin consider a disk, which due to the self-regulation 
has a Toomre parameter $Q \simeq 1$. It is accreting at a constant rate 
equal to the accretion rate within the parent collapsing core equal to 
$\epsilon C_s^3 / G$ (Shu 1977) where $\epsilon$ is a factor of a few.
Assuming stationarity, this accretion rate must be equal to 
the viscous rate 
 given by eq.~(\ref{dM_dt}) leading to 
$ \dot{M} = 3 \pi \alpha \Sigma c_s^2 / \Omega = \epsilon C_s^3 / G$.
Assuming that the disk is marginally critical, that is the 
criteria of Gammie (2001) is just satisfied, they 
estimate the $\alpha$ viscosity parameter to be about 0.23. 

With eq.~(\ref{heating_kep}), the heating rate, $F_v$, due 
to the viscous dissipation in the disk is thus 
known. The cooling rate per face is assumed to be
the classical black body radiation 
$F_r = (16 / 3) \sigma T^4 / (\Sigma \kappa)$,
where $\kappa \simeq \kappa_R^0 T^2$ is the Rosseland 
opacity taken from Semenov et al. (2003) and estimated 
to be $\kappa _R ^0 = 3 \times 10^{-4}$ cm$^{2}$g$^{-1}$.
The thermal balance leads to the relation $F_r = F_v$. 

Together with the conditions $Q >  1$,  the expression of the flux
$\dot{M}=\epsilon C_s^3 / G$ and the thermal balance lead to an 
estimate for the critical value of the rotation
frequency, $\Omega _{crit}$ above which the disk is stable
\begin{eqnarray}
\Omega > \Omega _{crit} \simeq 3.6 \left( { G^2 \mu^2 \sigma\over \alpha \kappa _R^0 k_B^2}
 \right)^{1/3} \simeq 4.3 \times 10^{-10} {\rm s}^{-1}, 
\label{omega_crit}
\end{eqnarray}
where the relation $c_s^2 = k_b T/ \mu$ has been used.

For a Keplerian disk, we have $\Omega^2= G M / R^3$, which leads to
\begin{eqnarray}
\Omega \simeq 2 \times 10^{-10} \left( { M \over 1 M_\odot} \right)^{1/2}
\left( { R \over 100 {\rm AU}} \right)^{-3/2} {\rm s}^{-1}.
\end{eqnarray}
The combination with eq.~(\ref{omega_crit}) shows that typically 
fragmentation can occur in mechanically heated disks, for radius 
larger than $\simeq 60$ AU. This offers an appealing explanation
for the brown dwarf desert (McCarthy \& Zuckerman 2004), i.e. the 
marked dearth for close brown dwarfs companions around stars.
In these conditions, the typical temperature within the disk
is about 15 K while the typical fragment mass is a few Jupiter mass.

Whitworth \& Stamatellos (2006) arrive to similar conclusions using related 
although slightly different arguments. First, they do not assume that 
the disk is mechanically heated but rather heated from the star radiation.
Second, instead of using the  Gammie criterion, they follow more 
closely the derivation presented in section 2.1.3 writing that the mechanical 
$PdV$ heating during contraction must be compensated by the cooling.
They estimate a slightly larger radii for the fragmentation to occur
and also propose that the brown dwarf desert can be naturally explained
in this context as follows (e.g.
 Stamatellos \& Whitworth 2009, figure 6).
Fragments that form in the disk (around
70-100 AU) migrate into the inner disk region and grow in mass fast to
become low-mass stars (they start from 1-3 Jupiter mass and grow to 0.2-0.3
solar mass) 
while  fragments that form further out, tend to migrate in the inner disc
region as well but they endure 3-body interactions with the most 
massive fragments
(and the central star) which are there. They are thus thrown out back to
outer disk region and the region close to the central star is kept clear
of brown dwarfs.

\subsubsection{The role of irradiation}
\label{irradiation}
The heating due to the central star is likely 
important although still controversial. An estimate for the temperature
is simply given by the radiative equilibrium
\begin{eqnarray}
\sigma T^4 = { L _* \over 4 \pi R_d^2}.
\end{eqnarray}
which leads to 
\begin{eqnarray}
 T = { L _*^{1/4} \over (4 \pi \sigma)^{1/4} R_d^{1/2} } \simeq 300 K \left( { L_* \over L_\odot} \right)^{1/4}
\left( { R_d \over 1 {\rm AU}} \right)^{-1/2}.
\end{eqnarray}
This shows that the temperature is typically larger that when the disk is heated 
by mechanical dissipation only for radius of the order of 100 AU.
Thus, the radius of the brown dwarf desert estimated by Whitworth \& Stamatellos (2006) is 
slightly larger and about $\simeq$150 AU.

However, the luminosity is not only due to the stellar luminosity but also to the 
accretion luminosity.
For a fiducial  one solar mass star accreting at a rate of about 
$10^{-5}$ M$_\odot$ s$^{-1}$, the accretion luminosity  is about
\begin{eqnarray}
L_{\rm acc} = {G M_* \over R_*} \dot{M} \simeq 150 L_\odot \left( { R_* \over R_\odot} \right)^{-1}
{M_* \over 1 M_\odot} {\dot{M} \over 10^{-5} {\rm M_\odot yr^{-1}}},
\end{eqnarray}
which is much larger than the stellar luminosity. Consequently, as noted by 
Matzner \& Levin (2005) and confirmed by the numerical simulations performed by 
Bate (2009) and Offner et al. (2009), radiation substantially reduces the fragmentation 
of self-gravitating disks. Indeed Matzner \& Levin (2005) even conclude
that it may entirely suppress the formation of brown dwarfs by 
disk fragmentation.

At this stage, another complication must be taken into account. The accretion 
may not be continuous but instead could occur in a burst mode as 
 advocated by Vorobyov \& Basu (2006). In their case,  this is due to disk non-linear dynamics
which leads to the formation of small dense clumps within the disks that increase 
drastically the accretion when they fall into the central object.
If this is the case, the implication is that the heating due to the accretion 
luminosity may also be intermittent leaving the possibilities that fragmentation could 
occur inbetween two bursts of accretion.  Recent simulations have been performed 
by Stamatellos et al. (2011b) along this line.  They indeed conclude that while disk 
fragmentation 
does not happen when the accretion is assumed to be constant, it does happen 
between the burst of accretion when it is irregular. 

To conclude, it seems that the possibilities of 
brown dwarfs being formed by disk fragmentation is eventually linked to 
how accretion proceeds within the disk and  to a large extent remains a matter 
of debate.

\subsubsection{The role of the magnetic field}
So far, the impact of the magnetic field has not been considered while 
as emphasized in several studies, it may have a crucial 
impact on disk formation and evolution. Here we simply summarize 
the most important ideas without going into the details.

$\bullet$ Magnetic braking \\
Due to the generation of torsional Alfv\'en waves which propagate and transfer 
angular momentum from the cloud to the intercloud medium (Mouschovias \& Paleologou 1979, 
Shu et al. 1987),
 the magnetic field is able to efficiently brake interstellar clouds. 
To estimate the time scale 
over which this process is occurring, let us consider an intercloud medium 
of density $\rho_{\rm icm}$ and let us assume that the magnetic field is parallel to 
the rotation axis. The waves propagate  at the Alfv\'en speed, 
$V_a= B / \sqrt{4 \pi \rho_{\rm icm}}$ along a cylinder parallel to the magnetic field. 
Significant braking will arise when the 
waves have transmitted to the intercloud medium a substantial fraction of the cloud angular 
momentum. This is the case, when the waves have reached a distance from the cloud, $l$, 
such that $l \times \rho_{\rm icm} \simeq R \times \rho_0$. That is to say the waves 
have been able to transfer angular momentum to a mass of intercloud medium comparable
to the mass of the cloud. This gives an estimate for the magnetic braking time,
in case where the magnetic field and the rotation axis are aligned:
\begin{eqnarray}
\tau _{\rm br} \simeq {R \over V_a} {\rho_0 \over \rho_{\rm icm}}. 
\label{brake}
\end{eqnarray}
The braking time increases when  $\rho_ {\rm icm}$ decreases because if the intercloud 
medium has a low inertia, its angular momentum is small. \\

$\bullet$ The magnetic braking catastrophe \\
Thanks to the progress of numerical schemes and computers, 
numerical calculations  of the collapse of magnetized cores have been recently
performed. One of the conclusions shared by several authors (e.g. Allen et al. 2003, 
Price \& Bate 2007, Mellon \& Li 2008, Hennebelle \& Fromang 2008) is that 
the magnetic field has a very severe impact on disk formation. Indeed, 
these authors conclude that even for relatively weak fields, the 
 formation of big massive disks that form in the hydrodynamical case, 
can be suppressed at least in the early phase of the collapse though 
Hennebelle \& Ciardi (2009) found that the problem is less severe when the 
magnetic field is inclined with respect to the rotation axis. 
This is because the magnetic field is strongly amplified during the collapse and 
transports very efficiently the angular momentum (e.g. Galli et al. 2006). 
This behaviour has been found to persist when non-ideal MHD is taken into 
account (e.g. Mellon \& Li, 2009 for ambipolar diffusion) though small scale disks 
can form at a scale of a few tens of AU due to ohmic dissipation at high densities
(Dapp \& Basu 2011, Machida et al. 2011).
These  models have received some support from the  comparisons between 
high resolution observations and synthetic ones performed using various models
(Maury et al. 2010). Observational data are in better agreement with magnetized
models than with purely hydrodynamical ones. The main difference comes
from the presence of a big massive disk in the latter case. 
An alternative view has been proposed by Stamatellos et al. (2011a) who argue
that the massive  disks do not live long because they quickly fragment, thus the 
chance to see them is low. \\

$\bullet$ Disk stabilization by magnetic field line twisting \\
When the magnetic field is sufficiently low, the magnetic braking
is not strong enough to remove efficiently the angular momentum 
and disks form. However, it has been found that fragmentation 
is much reduced if not suppressed (Hennebelle \& Teyssier 2008, 
Machida et al. 2008, Duffin \& Pudritz 2009). 
This is due to the twisting of the 
magnetic field lines by the Keplerian velocity profile that 
quickly triggers the growth of a toroidal component. 
The latter tends to stabilize the disk since it operates,
in the loosely sense, as an effective pressure that adds up 
to the thermal pressure. Note that at this stage, the impact 
of non-ideal MHD has not been investigated. This conclusion 
could change, if the field turns out to be not sufficiently coupled to the 
gas in the disk. \\ \\

The general trend is therefore that the magnetic field tends
to quench the formation of brown dwarfs by disk fragmentation 
because  it tends to suppress the formation of big massive 
disks at the class 0 phase (leaving probably small tens of AU disk) and 
whenever they form, magnetic field tends to stabilize them. The final 
answer will depend on the distribution of magnetic intensities in 
dense cores and also on the non-ideal MHD processes.

\section{Statistics from large scale numerical simulations}

The approaches developed in previous sections attempted to identify 
a particular situation responsible of the formation of low-mass stars and brown 
dwarfs. More specifically, the gravo-turbulent theory 
emphasizes the role of turbulent density fluctuations while the fragmentation
disk theory emphasizes the role of disk instabilities.
In reality, both processes should be considered simultaneously and treated 
as a continuum. In particular, the impact of rotation and gravitational 
density enhancement which are at the very heart of the disk fragmentation
theory should be taken into account in the gravo-turbulent theory. 
Large scale numerical simulations have been performed along this line
 using either  smoothed particle hydrodynamics or adaptive mesh refinement
techniques. 
This approach, which avoids many of the assumptions made in the previous sections, 
has the advantage of providing direct statistics which can be compared 
with the observational data. Given their complexity, it is however 
not always easy ${\it i)}$ to check the reliability of these 
predictions (numerical convergence and sensibility to the physical processes)
and ${\it ii)}$ to interpret the physical mechanisms at play. 

\subsection{Numerical and physical setup}
The typical setup used in this kind of calculations consists
in a cloud of hundreds of solar masses (e.g. Bate et al. 2003). The shape is 
most of the time spherical  and the density profile goes from a 
uniform density to an $r^{-2}$ dependence (Girichidis et al. 2011).
The problem is largely determined by the values of the
 thermal and turbulent energies which are typically a
 few percent or less for the former and a factor of a few for the 
latter. A ${\it turbulent}$ velocity field, which has a powerspectrum
close to the Kolmogorov one and random phases, is initially setup.
Finally, sink particles that mimic the formation of stars, are usually used.
Such Lagrangian entities (e.g. Federrath et al. 2010b) can accrete 
the surrounding gas and interact with it through  gravity.
To mimic the dust opacity, a barotropic equation of state is usually employed
leading for example to $c_s^2 = c_{s,0}^2 (1 + (n / n_c)^{2/3})$
with $n_c \simeq 10^{10}$ cm$^{-3}$ (note that the exponent $2/3=5/3-1$ is valid 
until the internal rotation level of H$_2$ are not excited after which it 
should typically be equal to $2/5=7/5-1$). As described below, calculations 
including  radiative transfer and heating by the  stars 
have been performed as well though due to 
very large computing overhead, the statistics obtained so far are only preliminary.
The same is true for the magnetic field. Simulations of magnetized massive clumps 
(Price \& Bate 2009, Peters et al. 2010, Hennebelle et al. 2011, Seifried et al. 2011,
Commer{\c c}on et al. 2011) have attempted to demonstrate trends rather than to obtain 
statistics. The general conclusion matches the trends inferred from 
the core collapse calculation presented above,  magnetic field tends to reduce 
the fragmentation of these massive clouds.

\subsection{Two new mechanisms for the formation of brown dwarfs}

From the results of large scale numerical simulations, two 
new mechanisms for the formation of brown dwarfs have been inferred. 
The first one is the formation of brown dwarfs by ejection and 
the second one is the formation within dense filaments induced 
by gravity during collapse. Both mechanisms rely on 
 strong density enhancements triggered by gravity.

\subsubsection{Formation of brown dwarfs by ejection}
In  collapsing massive, turbulent clumps, massive centrifugally supported disks
form as turbulence provides angular momentum. These disks 
 quickly fragment as described above (at least when radiative transfer and magnetic 
field are ignored). As usually a small cluster of objects formed, they undergo 
N-body interactions and the less massive objects can easily be ejected from the 
parent cloud when the velocity they acquire during a close encounter is large 
enough. When they leave the cloud, there is no more gas around, that they 
could accrete and therefore their masses do not increase along time as it is the 
case for an object embedded in the parent clump. This mechanism 
has  originally been proposed by Watkins et al. (1998) and Reipurth \& Clarke (2001)
and later observed and quantified in large scale simulations (Bate et al. 2002, Bate 2012).

\subsubsection{Formation of brown dwarfs in dense filaments}
The physics of this mechanism is simply that as the clump  collapses, 
dense filaments form by a combination of shocks and local gravitational 
collapse (which in particular amplifies the anisotropies and tend to form 
more filaments). Within these dense structures, the density is larger and the Jeans 
mass is smaller, thus smaller objects including brown dwarfs can form. Note that while the 
preceding mechanisms was essentially a formation by disk fragmentation, the present one 
is very similar to the gravo-turbulent mechanism presented previously. The main difference
is that the PDF instead of being due to turbulent compression, is due to gravitational
contraction. This mechanism has been studied in detail by Bonnell et al. (2008). They 
show for example that there is a correlation between the initial Jeans mass and the 
final mass of the stars, suggesting that indeed this mechanism is at play. 

Bate et al. (2002) estimate that in their barotropic calculations, the first 
mechanism is dominating over the second by a factor of the order of $\simeq$3.
However, as the fragmentation of disk is much affected by 
the heating from the central star and the magnetic field, it is unclear 
whether this conclusion is really robust.

\subsection{Impact of the radiative transfer}
\label{radiative}
Before discussing some of the quantitative results obtained 
in large scale numerical simulations, it is useful to 
discuss qualitatively the influence of the radiative feedback further. 
Indeed, it has been found to play an important role in reducing 
the fragmentation by Krumholz et al. (2007), Bate (2009) and
Commer{\c c}on et al. (2011). Generally speaking the reason 
is similar to what is discussed in section~\ref{irradiation}, that 
is to say around the accreting protostars, the gas is heated 
due to the intense accretion luminosity and thus the temperature 
is higher than what is typically obtained by using a barotropic 
equation of state. Although no simple estimate  
of the efficiency in reducing the fragmentation  has been 
provided, 
  this effect is found to be substancial 
in the above mentioned studies. For example, Bate (2009) concludes
that the number of brown dwarfs may be lower by a factor of about 
$\simeq 3$ in his radiative calculations. \\

Another possible consequence, though not firmly established yet, is that 
the universality of the peak of the IMF could be due to the heating by the 
stars themselves. 
This is based on a scaling argument proposed by Bate (2009). 
The argument is as follows.

In the optically thick regime, the temperature around a
star is given by $L_* \simeq 4 \pi \sigma r^2 T^4$ while 
a Jeans length in this region is given by the relation
$c_s^2 = k T / m_p \simeq G M / \lambda_J$ that is to say 
the thermal support is roughly equal to the gravitational 
energy for a mass $M \simeq (4 \pi / 3) \rho \lambda_J^3$.
This leads to the relation
\begin{eqnarray}
\lambda_J^{5/2} \simeq {3 k L_*^{1/4} \over (4 \pi)^{5/4} G m_p \sigma^{1/4} } \rho^{-1}.
\end{eqnarray}
The Jeans mass which is simply estimated as $M_J \simeq (4 \pi/ 3) \rho \lambda_J^3$
is then found to be
\begin{eqnarray}
M_J =  (4 \pi / 3)^{-1/5} (k/G m_p)^{6/5} L_*^{1/10} (4 \pi \sigma)^{-1/10} \rho^{-1/5}.  
\label{mass_jeans_heating}
\end{eqnarray}
Thus instead of getting a density dependence $\rho^{-1/2}$ as in the 
isothermal case, one gets $M_J \propto \rho^{-1/5}$ which is 
much shallower dependence seemingly leading to an IMF which 
is less dependent on the cloud parameters. 
Although it constitutes an interesting idea, various aspects of the scheme 
need further explanations. For example, the first generation of stars 
that formed in a molecular cloud, is not affected by this heating and therefore
would have different characteristics unless this effect starts operating while 
the stars are still forming (for example by modifying the Jeans mass that would 
otherwise be accreted by the central star). Another question is the fraction 
of the gas in the molecular cloud which is affected by this heating. 
As this effect strongly depends on the distance to the stars, the pieces of gas 
which are too far from stars are not heated significantly.

\subsection{The IMF and its sensibility to the initial conditions}

Numerical simulations performed with SPH (e.g. Jappsen et al. 2005, 
Smith et al. 2008, Bate 2009, 2012) or using AMR and sink particles 
(e.g. Girichidis et al. 2011) tend to produce distributions which 
present similarities with the observed IMF, that is to say 
they present a powerlaw at high masses, which has an index
compatible with the Salpeter value and a peak at some smaller mass.
While the index of the high mass part appears to be seemingly robust (although
the distributions are not always clearly powerlaws), 
the position of the peak  vary
as expected with the initial conditions (for example Bate \& Bonnell 2005 
show that it depends on the thermal energy) and the equation 
of state  (e.g. Jappsen et al. 2005). The dependence on the 
initial density profiles found by Girichidis et al. (2011) is 
very drastic. For example, while when  they start with a uniform density 
profile, they obtain the formation of many objects that mimics the IMF (though 
peaking at too small mass), 
they find in the case of an $r^{-2}$ profile, that a single object can be obtained.
In general, these simulations tend to find that the peak of the mass distribution
is too small compared to the peak of the IMF (which is typically equal to 0.3 $M_\odot$).
For example Bate et al. (2003) conclude that too many brown dwarfs (typically a factor 3) 
form in his simulations. 

Interestingly, Bate (2012) includes radiative transfer as described 
in \S~\ref{radiative} and finds that the fragmentation is significantly reduced 
 with respect to his former simulations (Bate et al. 2003) and obtains a 
mass distribution very similar to the IMF. Note that at this stage, this 
result is obtained for one particular clump having a specific initial 
density and turbulence strength. One of the remaining question 
 is therefore what determines the initial cloud parameters (mean density, 
mass and Mach number) which all influence the mass distribution. 
As  argued by Bate (2009)   
when  radiation is included, the mass distribution may depend 
only weakly on, at least, the initial density though this should 
clearly be demonstrated by running simulations with different initial conditions and high 
resolution.
Another word of caution is the influence of the magnetic field, not treated in these
work. As advocated in Commer{\c c}on et al. (2010, 2011), the magnetic braking has 
a drastic influence on  the radiative feedback by 
transporting angular momentum away and leading to enhanced local accretion rates.
It is also difficult to properly simulate as it requires a high numerical resolution.

\subsection{Physical origin of the IMF in numerical simulations}

The mechanism responsible for setting the IMF in the above mentioned numerical 
simulations, is not easy to identify and remains a matter of debate. 

Bate (2012) argues that it is due to competitive accretion. His main argument 
is that the more massive stars at the end of the simulations are the ones, 
which have accreted longer. Moreover, the accretion often ends after 
the star undergoes a dynamical interaction and is therefore ejected from 
the densest part of the parent cloud. Let us stress that  
the first part of this argument is also compatible with the massive stars 
having a larger accretion reservoir available as suggested by 
Hennebelle \& Chabrier (2008). This is because bigger cores require
more time to collapse. An important test to distinguish between the various 
scenarii would be to quantify the spatial distribution of the gas accreted 
by the sink.

On the other hand, the SPH simulations performed by Smith et al. (2008) show a 
clear correlation between the initial masses (see also Bonnell et al. 2008) 
within the gravitational well and 
the final sink masses up to a few local freefall times (see Chabrier \& Hennebelle 2010 for
a quantitative analysis),  suggesting that 
the initial prestellar cores do not fragment into many objects. 
As time goes on, the correlation
becomes weaker but seems to persist up to the end of their run. Massive stars, 
on the other hand, 
are weakly correlated with the mass of the potential well in which they form.
Whether their mass was contained into a larger more massive well 
with which the final sink mass would be well correlated  remains an open issue, 
which needs to be further investigated.

Finally, we reiterate that  the magnetic field (e.g. Machida et al. 2005, 
Hennebelle \& Teyssier 2008, Price \& Bate 2009, Peters et al. 2010, Hennebelle et al. 2011) 
and even more the combination between magnetic field and radiative feedback 
(Commer{\c c}on et al. 2011) seem to reduce 
the fragmentation, suggesting that the core-sink correlation found in Smith et al. (2008) 
should improve if such processes were included.
Clearly, these questions require careful investigations.

\section{Conclusion}
We have reviewed the most recent theories which have been proposed to explain the 
origin of the formation of low-mass stars and brown dwarfs.
First, we have described the physical processes at play in star 
formation as well as the most fundamental results. 
Second, we have presented the theories which attempt to explain the origin 
of the IMF.
Two main categories received particular attention: the theories
based on accretion and the ones based explicitly on turbulence. It should be 
stressed that these theories are not all exclusive of each other  and may
apply in different ranges of mass. For instance, the turbulent dispersion theory
 calculates the distribution of the initial mass accretion 
reservoirs;  it is not incompatible with the stopped accretion theories and 
with the competitive accretion as long as mass redistribution/competition  occurs within one 
parent core reservoir. 
The question as to whether one of these mechanisms is dominant is yet unsettled. 
Detailed comparisons between systematic sets of simulations, as done in Schmidt et al. (2010), 
or observations, and the various analytical 
predictions is clearly mandatory to make further progress.
Third, we have described the theory for brown dwarf formation based 
on massive disk fragmentation. While it could in principle lead to the formation of 
at least a fraction of the brown dwarfs, large uncertainties remain regarding the role 
of magnetic field and radiation that could drastically reduce the importance of this 
mechanism. Finally, we described the large numerical simulations, which have been 
performed to study massive clump collapse and fragmentation. The introduction of 
magnetic field and radiative feedback has been a significant improvement of the last 
years, which unsurprisingly has revealed that both have a major impact on the
outcome of star formation. One important challenge for the next decades is certainly 
to perform large scale simulations which include both effects, including non-ideal MHD, 
with a sufficient resolution.  \\ \\

{\bf Acknowledgments}
It is a pleasure to thank Corinne Charbonnel, C\'ecile Reyl\'e and Mathias Schultheis for the organization of 
a very nice school. I also thank Dimitri Stamatellos, Farzana Meru, Christopher Matzner and
 S\'ebastien Fromang for enlighting discussions. I thank Benoit Commer{\c c}on and Andrea Ciardi 
for a critical reading of the manuscript.


\begin{thebibliography}{}



\bibitem[]{adams95} {Adams, F., \& Fatuzzo, M.} 1996, \textit{ApJ}, 464, 256



\bibitem{allen}
Allen, A., Li, Z.-Y., Shu, F.,
{\ 2003, \textit{ApJ}, 599, 363}


\bibitem[]{andre07} Andr\'e, Ph., Belloche, A., Motte, F. \& Peretto, N., 2007, \textit{A\&A},  472, 519

\bibitem[]{andre109} Andr\'e, Ph., Basu, S. \& Inutsuka, S-I., 2009, in  \textit{Structure Formation in Astrophysics}, G. Chabrier Ed., Cambridge University Press

\bibitem[]{andre10} Andr\'e, P., Men'shchikov, A., Bontemps, S. et al.
\textit{A\&A}, 2010, 518L, 102



\bibitem{balbus}
Balbus, S., Hawley, J., 
{\ 1998, Rev. Mod. Phys., 70, 1}

\bibitem[]{bastian}
Bastian, N., Covey, K., Meyer, M., 2010, \textit{ARA\&A}, 48, 339



\bibitem{basu1} Basu, S., Ciolek, G. 2004, \textit{ApJ}, 607, L39 

\bibitem[]{BasuJones04} Basu, S., Jones, C., 2004, \textit{MNRAS}, 347, L47 

\bibitem{basu2} Basu, S., Mouschovias, T. 1995, \textit{ApJ}, 452, 386

\bibitem[]{bate02} Bate, M., Bonnell, I., Bromm, V., 2002, \textit{MNRAS}, 332L, 65

\bibitem[]{bate03} Bate, M., Bonnell, I., Bromm, V., 2003, \textit{MNRAS}, 339, 577

\bibitem[]{bate09} Bate, M., 2009, \textit{MNRAS}, 392, 1363

\bibitem[]{bate12} Bate, M., 2012, \textit{MNRAS}, 419, 3115

\bibitem[]{BateBonnell05} Bate, M., Bonnell, I., 2005, \textit{MNRAS}, 356, 1201


\bibitem[]{binney}
Binney, J., Tremaine, S., 1987, Galactic Dynamics, gady.book

\bibitem[]{bonnel01}
Bonnell I., Bate M., Clarke C., Pringle J., 2001, \textit{MNRAS} 323, 785

\bibitem[]{bonnel08}
Bonnell I., Clark, P., Bate M., 2008, \textit{MNRAS} 389, 1556


\bibitem{bonnor} Bonnor, W. 1956, \textit{MNRAS}, 116, 351

\bibitem[]{C03a} Chabrier, G., 2003, \textit{PASP}, 115, 763

\bibitem[]{ch10} Chabrier, G.,  Hennebelle, P., 2010, \textit{ApJL}, 725L, 79

\bibitem[]{ch11} Chabrier, G.,  Hennebelle, P., 2011, \textit{A\&A}, 534, 106


\bibitem[]{Clark2007} Clark, P., Klessen, R., Bonnell, I., 2007,  \textit{MNRAS}  379, 57

\bibitem{commercon2}
Commer{\c c}on, B., Hennebelle, P., Audit, E., Chabrier, G., Teyssier, R.,
{\ 2010, \textit{A\&A}, 510L, 3}

\bibitem[]{commer2011} Commer{\c c}on, B., Hennebelle, P., Henning, T., 
2011, \textit{ApJ}, 742L, 9

\bibitem[]{dapp} Dapp, W., Basu, S., 2011, \textit{A\&A}, 521L, 56

\bibitem{duffin}
Duffin, D., Pudritz, R., 
{\ 2009, \textit{ApJ}, 706L, 46}

\bibitem[]{duquennoy} Duquennoy, A., Mayor, M., 1991, \textit{A\&A}, 248, 485

\bibitem[]{dutrey} Dutrey, A., Duvert, G., Castets, A., Langer, W., Bally, J., 
Wilson, R.,  1991, A\&A, 247L, 9




\bibitem[]{Elmegreen1}
Elmegreen B.G.,  1997, \textit{ApJ} 486, 944


\bibitem[]{elmegreen08} Elmegreen, B., Klessen, R., Wilson, C., 2008, \textit{ApJ}, 681, 365

\bibitem[]{federrath10} Federrath, C., Roman-Duval, J., Klessen, R., Schmidt, W., Mac Low, M.-M.,
2010a, \textit{A\&A}, 512, 81

\bibitem[]{federrath10} Federrath, C., Banerjee, R., Clark, P., Klessen, R.,
2010b, \textit{ApJ}, 713, 269

\bibitem{fiege} Fiege, J., Pudritz, R. 2000, \textit{MNRAS}, 311, 105

\bibitem{foster} Foster, J., Chevalier, R. 1993, \textit{ApJ}, 416, 303 

\bibitem{galli} Galli, D., Lizano, S., Shu, F., Allen, A., 2006, \textit{ApJ}, 647, 374

\bibitem{gammie} Gammie, C., 2001, \textit{ApJ}, 553, 174

\bibitem{girichidis}
Girichidis, P., Federrath, C., Banerjee, R., Klessen, R.,
{\ 2011, \textit{MNRAS}, 413, 2741}


\bibitem{goldbaum} Goldbaum, N., Krumholz, M., Matzner, C., McKee, C., 
2011, \textit{ApJ}, 738, 101

\bibitem[]{haisch} Haisch, K., Lada, E., Lada, C., 
{2001, \textit{ApJ}, 553L, 153} 

\bibitem[]{hartmann} Hartmann, L., 
{2009, apsf.book}

\bibitem[heiderman]{heiderman10} 
Heiderman, A., Evans, N., Allen, L., Huard, T., Heyer, M., 
{\ 2010, \textit{ApJ}, 723, 1019}

\bibitem[]{heithausen}
Heithausen, A., Bensch, F., Stutzki, J., Falgarone, F., \& Panis, J.-F.,
 1998, \textit{A\&A}, 331, L65

\bibitem[Hennebelle et al. (2003)]{hen03} 
Hennebelle, P., Whitworth, A., Gladwin, P., Andr\'e, Ph. 2003, \textit{MNRAS}, 340, 870


\bibitem[Hennebelle \& Fromang 2008]{HF08} Hennebelle, P., Fromang, S., 2008, \textit{A\&A}, 477, 9 

\bibitem[Hennebelle \& Teyssier 2008]{HT08} Hennebelle, P., Teyssier, R., 2008, \textit{A\&A}, 477, 25

\bibitem[Hennebelle  ciardi 2009]{HC09} Hennebelle, P., Ciardi, A., 2009, \textit{A\&A}, 506L, 29



\bibitem[]{HC08a} Hennebelle, P., Banerjee, R., V\'azquez-Semadeni, E., Klessen, R. 
\& Audit, E., 2008, \textit{A\&A}, 446, 43

\bibitem[]{HC08} Hennebelle, P., Chabrier, G., 2008, \textit{ApJ}, 684, 395 

\bibitem[]{HC09} Hennebelle, P.,  Chabrier, G., 2009, \textit{ApJ}, 702, 1428

\bibitem[]{HC11} Hennebelle, P.,  Chabrier, G., 2011, \textit{ApJ}, 743L, 29

\bibitem[]{Hetal11} Hennebelle, P., Commer{\c c}on, B., Joos, M., et al. 
2011, \textit{A\&A}, 528, 72

\bibitem{hopkins} Hopkins, P., 2012, arXiv1201.4387

\bibitem{hoyle} Hoyle, F. 1953, \textit{ApJ}, 118, 513 

\bibitem[jappsen 2006]{jappsen} Jappsen, A., Klessen, R., Larson, R., Li, Y., Mac Low,M.-M., 2005, \textit{A\&A}, 435, 611   

\bibitem{Jeans} Jeans, J. 1905,  \textit{ApJ}, 22, 93

\bibitem{klessen} Klessen, R., Hennebelle, P., 2010, \textit{A\&A}, 520, 17

\bibitem[Klessen et al. 2000]{KHML00}
Klessen, R., Heitsch, F., Mac Low, M.-M. 2000, \textit{ApJ}, 535, 887



\bibitem[]{Kritsuk_etal07}
Kritsuk, A. G., Norman, M. L., Padoan, P., \& Wagner, R. 2007, \textit{ApJ},
665, 416

\bibitem[]{kroupa}
Kroupa P.,  2002, \textit{Sci.} 295, 82

\bibitem[]{krumholz}
Krumholz M., \& McKee C.,  2005, \textit{ApJ}, 630, 250

\bibitem{krumholz3}
Krumholz, M.,  Klein, R., McKee, C.,
{2007, \textit{ApJ}, 656, 959}


\bibitem[]{kunz}
Kunz M., \& Mouschovias T.,  2009, \textit{MNRAS} 399L, 94



\bibitem{larson} Larson, R. 1969, \textit{MNRAS}, 145, 405


\bibitem[]{Larson92} Larson R., 1973, \textit{MNRAS}, 161, 133

\bibitem[]{Larson81} 
Larson R.,  1981, \textit{MNRAS}, 194, 809


\bibitem[]{Larson85} Larson R., 1985, \textit{MNRAS}, 214, 379



\bibitem{lequeux} Lequeux, J. 2005, in The interstellar medium, J. Lequeux (ed), EDP Sciences (Berlin: Springer)


\bibitem{li} Li, Z.-Y., Nakamura, F. 2005, \textit{ApJ}, 631, 411

\bibitem[]{lodato} Lodato, G., Rice, K., 2004, \textit{MNRAS},351, 630


\bibitem[]{mccarthy} McCarthy, C., Zuckerman, B., 2004, \textit{AJ}, 127, 2871

\bibitem[]{machida} Machida, M., Matsumoto, T., Hanawa, T., Tomisaka, K., 
2005, \textit{MNRAS}, 362, 382


\bibitem[]{machida1} Machida, M., Tomisaka, K.,  Matsumoto, T., Inutsuka, S.-i
, 2008, \textit{ApJ}, 677, 327


\bibitem[]{machida2} Machida, M., Inutsuka, S.-i, Matsumoto, T., 
2011, \textit{PASJ}, 63, 555


\bibitem[]{mackee2}
McKee C.F., \& Ostriker E.,  2007, \textit{ARA\&A}, 45, 565



\bibitem[]{mackee3}
McKee C.F., \& Tan J.C.,  2003, \textit{ApJ} 585, 850

\bibitem[]{MacLowKlessen04} MacLow, M.-M., \&  Klessen, R., 2004, \textit{Rev. Mod. Phys.}, 76, 125

\bibitem[]{masu} Masunaga, H., Inutsuka, S.-i., 2000, \textit{ApJ}, 531, 350

\bibitem[]{matzner1} Matzner, C., 2002, \textit{ApJ}, 566, 302


\bibitem[]{matzner3} Matzner, C., Levin, Y., 2005, \textit{ApJ}, 628, 817

\bibitem[]{matzner2} Matzner, C., 2007, \textit{ApJ}, 659, 1394

\bibitem{Maury} Maury, A., Andr\'e, P., Hennebelle, P., et al., 2010, \textit{A\&A}, 512, 40

\bibitem{mellon} Mellon, R., Li, Z.-Y., 2008, \textit{ApJ}, 681, 1356

\bibitem{mellon} Mellon, R., Li, Z.-Y., 2009, \textit{ApJ}, 698, 922


\bibitem{meru} Meru, F., Bate, M., 2011, \textit{MNRAS}, 411L, 1

\bibitem[]{mihalas} Mihalas, D., Mihalas, B., 1984, Foundations of radiation hydrodynamics, oup book





\bibitem{mouscho-spitzer} Mouschovias, T., Spitzer, L. 1976, \textit{ApJ}, 210, 326

\bibitem{mouschovias} Mouschovias, T., Paleologou, E. 1979, \textit{ApJ}, 230, 204

\bibitem{mouschovias} Mouschovias, T., Paleologou, E. 1981, \textit{ApJ}, 246, 48

\bibitem[]{myers2}
Myers P.C., 2009, \textit{ApJ}, 706, 1341


\bibitem[]{nagai}
Nagai, T., Inutsuka, S.-i., Miyama, S., 1998, \textit{ApJ}, 506, 306 

\bibitem[]{offner} Offner, S., Klein, R., McKee, C., Krumholz, M.,  
2009, \textit{ApJ}, 703, 131



\bibitem{ostriker} Ostriker, J. 1964 , \textit{ApJ}, 140, 1056

\bibitem{paczynski} Paczynski, B., 1978, Acta Astron., 28, 91

\bibitem[Padoan \& Nordlund 1997]{PadoanNordlund97} Padoan, P., Nordlund, A., \& Jones, B., 
1997, \textit{MNRAS}, 288, 145

\bibitem[]{PN99}
Padoan, P., \& Nordlund,  1999, \textit{ApJ}, 526, 279 

\bibitem[]{PN02}
Padoan, P., \& Nordlund,  2002, \textit{ApJ}, 576, 870


\bibitem[Padoan-Nordlund]{PadoanNordlund11} 
Padoan, P., Nordlund, A., 
{\ 2011, \textit{ApJ}, 730, 40}


\bibitem{penston} Penston, M. 1969, \textit{MNRAS}, 144, 425

\bibitem{peters}
Peters, T., Banerjee, R., Klessen, R., MacLow, M.-M., Galv\'an-Madrid, R., Keto, E., 
{\ 2010, \textit{ApJ}, 711, 1017}


\bibitem[]{PressSchechter74} Press, W., \& Schechter, P., 1974, \textit{ApJ}, 187, 425


\bibitem{price}
Price, D., Bate, M.,
{\ 2007, \textit{MNRAS}, 377, 77}


\bibitem{price} Price, D., Bate, M., 2009, \textit{MNRAS}, 398, 33  

\bibitem[]{price-podsi} Price, N., Podsiadlowski, P., 1995, \textit{MNRAS}, 275, 1041


\bibitem{pringle} Pringle, J., 1981, \textit{ARA\&A}, 19, 137

\bibitem[]{rees} Rees, M., 1976, \textit{MNRAS}, 176, 483

\bibitem[]{reipurth} Reipurth, B., Clarke, C., 2001, \textit{AJ}, 122, 432


\bibitem[Salpeter 1955]{S55} Salpeter, E., 1955, \textit{ApJ}, 121, 161

\bibitem[]{scalo}
Scalo J.,  1986, \textit{FCPh}, 11, 1

\bibitem[]{schmidt09}
Schmidt, W.,  Federrath, C., Hupp, M., Kern, S., Niemeyer, J., 
2009, \textit{A\&A}, 494, 127

\bibitem[]{schmidt10}
Schmidt, W., Kern, S., Federrath, C., Klessen, R., 
2010, \textit{A\&A}, 516, 25

\bibitem{seifried}
Seifried, D., Banerjee, R., Klessen, R., Duffin, D., Pudritz, R., 
2011, \textit{ApJ}, 417, 1054

\bibitem[]{semenov}
Semenov, D., Henning, Th., Helling, Ch., Ilgner, M., Sedlmayr, E., 
2003, \textit{A\&A}, 410, 611


\bibitem[]{S95} Silk, J., 1995,  \textit{ApJ}, 438, L41



\bibitem[]{shu77}
Shu F., 1977, \textit{ApJ}, 214, 488

\bibitem[Shu et al. 1987]{SAL87}
Shu, F. H., Adams, F. C., Lizano, S. 1987, \textit{ARA\&A} 25, 23



\bibitem{spitzer} Spitzer, L. 1942, ApJ, 95, 329

\bibitem[]{smith}
Smith, R., Clark, P., Bonnell, I.,
2008, \textit{MNRAS}, 391, 1091


\bibitem{stama} Stamatellos, D.,  Whitworth, A.,
2009, MNRAS, 392, 413

\bibitem{stama2} Stamatellos, D., Maury, A.,  Whitworth, A., Andr\'e, P.,
2011a, \textit{MNRAS}, 413, 1787

\bibitem{stama} Stamatellos, D., Hubber, D., Whitworth, A.,
2011b, \textit{ApJ}, 730, 32


\bibitem{terebey}
Terebey, S., Shu, F., Cassen, P., 1984, \textit{ApJ}, 286, 529

\bibitem[]{toomre}
Toomre, A., 1964, \textit{ApJ}, 139, 1217

\bibitem[]{troland1986}
Troland T., Heiles C., 
1986, \textit{ApJ}, 301, 339

\bibitem[]{vazquez1994}
V\'azquez-Semadeni, E., 
1994, \textit{ApJ}, 423, 681

\bibitem[]{vorobyov}
Vorobyov, E., Basu, S., 2006, \textit{ApJ}, 650, 956

\bibitem[{{Watson} {et~al.}(2007){Watson}, {Stapelfeldt}, {Wood}, \&
  {M{\'e}nard}}]{Watson07}
{Watson}, A.~M., {Stapelfeldt}, K.~R., {Wood}, K., \& {M{\'e}nard}, F. 2007,
  Protostars and Planets V, 523

\bibitem{watkins} Watkins, S., Boffin, H., Francis, N., Whitworth, A.,
1998, \textit{MNRAS}, 300, 1214

\bibitem[]{whitworth}
Whitworth, A., Bate, M., Nordlund, A., Reipurth, B., Zinnecker, H., 
2007, prpl.conf 459

\bibitem[]{whitworth}
Whitworth, A., Stamatellos, D., 
2006, \textit{A\&A}, 458, 817

\bibitem{whitworth} Whitworth, A., Summers, D., 1985, \textit{MNRAS}, 214, 1


\bibitem[]{zinnecker82}
Zinnecker H., 1982,  
in Glassgold A. E. et al., eds, Symposium on the Orion 
Nebula to Honour Henry Draper. New York Academy of Sciences, 
New York, p. 226 

\bibitem[]{zinnecker84}
Zinnecker H., 1984, \textit{MNRAS}, 210, 43  

\bibitem[1974]{zuc} Zuckerman, B., Evans, N. 1974, \textit{ApJ}, 192, L149.

\end{thebibliography}
\end{document}